\documentclass[12pt,preprint]{aastex}

\newcommand{\kms}{\ifmmode {\rm km\ s}^{-1} \else km s$^{-1}$\fi}
\newcommand{\Msun}{\ifmmode {\rm M}_{\odot} \else M$_{\odot}$\fi}
\newcommand{\Lsun}{\ifmmode {\rm L}_{\odot} \else L$_{\odot}$\fi}
\newcommand{\qo}{\ifmmode q_{\rm o} \else $q_{\rm o}$\fi}
\newcommand{\Ho}{\ifmmode H_{\rm o} \else $H_{\rm o}$\fi}
\newcommand{\ho}{\ifmmode h_{\rm o} \else $h_{\rm o}$\fi}
\newcommand{\ltsim}{\raisebox{-.5ex}{$\;\stackrel{<}{\sim}\;$}}

\newcommand{\vFWHM}{\ifmmode v_{\mbox{\tiny FWHM}} \else
                    $v_{\mbox{\tiny FWHM}}$\fi}
\newcommand{\CCF}{\ifmmode F_{\it CCF} \else $F_{\it CCF}$\fi}
\newcommand{\ACF}{\ifmmode F_{\it ACF} \else $F_{\it ACF}$\fi}
\newcommand{\Halpha}{\ifmmode {\rm H}\alpha \else H$\alpha$\fi}
\newcommand{\Hbeta}{\ifmmode {\rm H}\beta \else H$\beta$\fi}
\newcommand{\Hgamma}{\ifmmode {\rm H}\gamma \else H$\gamma$\fi}
\newcommand{\Hdelta}{\ifmmode {\rm H}\delta \else H$\delta$\fi}
\newcommand{\Lya}{\ifmmode {\rm Ly}\alpha \else Ly$\alpha$\fi}
\newcommand{\Lyb}{\ifmmode {\rm Ly}\beta \else Ly$\beta$\fi}
\newcommand{\HeI}{\ifmmode {\rm He}\,{\sc i}\,\lambda5876 \else 
	          He\,{\sc i}\,$\lambda5876$\fi}
\newcommand{\HeII}{\ifmmode {\rm He}\,{\sc ii}\,\lambda4686 \else 
	           He\,{\sc ii}\,$\lambda4686$\fi}

\newcommand{\ciii}{\ifmmode {\rm C}\,{\sc iii} \else C\,{\sc iii}\fi}

\newcommand{\oiii}{O\,{\sc iii}}

\shorttitle{The Black Hole Mass of NGC 4051}
\shortauthors{}

\received{}
\accepted{}

\begin{document}

\title{A Revised Broad-Line Region Radius and Black Hole Mass for the
Narrow-Line Seyfert 1 NGC~4051}

\author{ K.~D.~Denney\altaffilmark{1},
         L.~C.~Watson\altaffilmark{1},
         B.~M.~Peterson\altaffilmark{1,2},
         R.~W.~Pogge\altaffilmark{1,2}, D.~W.~Atlee\altaffilmark{1},
         M.~C.~Bentz\altaffilmark{1,3},
         J.~C.~Bird\altaffilmark{1},
         D.~J.~Brokofsky\altaffilmark{4,5},
         M.~L.~Comins\altaffilmark{1,6},
         M.~Dietrich\altaffilmark{1},
         V.~T.~Doroshenko\altaffilmark{7,8,16},
         J.~D.~Eastman\altaffilmark{1}, Y.~S.~Efimov\altaffilmark{8},
         C.~M.~Gaskell\altaffilmark{4,9},
         C.~H.~Hedrick\altaffilmark{4,6},
         S.~A.~Klimanov\altaffilmark{8,16},
         E.~S.~Klimek\altaffilmark{4,10},
         A.~K.~Kruse\altaffilmark{4}, J.~Lamb\altaffilmark{11},
         K.~Leighly\altaffilmark{12},
         T.~Minezaki\altaffilmark{13}, S.~V.~Nazarov\altaffilmark{8.16},
         E.~A.~Petersen\altaffilmark{4},
         P.~Peterson\altaffilmark{14},
         S.~Poindexter\altaffilmark{1}, Y.~Sakata\altaffilmark{15}
         K.~J.~Schlesinger\altaffilmark{1},
         S.~G.~Sergeev\altaffilmark{7,16},
         J.~J.~Tobin\altaffilmark{11}, C.~Unterborn\altaffilmark{1},
         M.~Vestergaard\altaffilmark{17,18},
         A.~E.~Watkins\altaffilmark{4}, and
         Y.~Yoshii\altaffilmark{13,19} }

\altaffiltext{1}{Department of Astronomy, 
		The Ohio State University, 
		140 West 18th Avenue, 
		Columbus, OH 43210; 
		denney, watson, peterson,
                pogge@astronomy.ohio-state.edu}

\altaffiltext{2}{Center for Cosmology and AstroParticle Physics, 
                 The Ohio State University,
		 191 West Woodruff Avenue, 
		 Columbus, OH 43210}

\altaffiltext{3}{Present address: 
		 Dept. of Physics and Astronomy,
		 4129 Frederick Reines Hall,
		 University of California at Irvine,
		 Irvine, CA 92697-4575;
		 mbentz@uci.edu}

\altaffiltext{4}{Department of Physics \& Astronomy, 
		 University of Nebraska, 
		 Lincoln, NE 68588-0111. }

\altaffiltext{5}{Deceased, Sept. 13, 2008}

\altaffiltext{6}{Present address: 
		 Astronomy and Astrophysics Department, 
		 Pennsylvania State University, 
		 525 Davey Laboratory, University Park, PA 16802}

\altaffiltext{7}{Crimean Laboratory of the Sternberg Astronomical Institute, 
	         p/o Nauchny, 98409 Crimea, Ukraine;
		 vdorosh@sai.crimea.ua}

\altaffiltext{8}{Crimean Astrophysical Observatory,
		 p/o Nauchny, 98409 Crimea, Ukraine;
		 sergeev, efim@crao.crimea.ua,
		 sergdave2004@mail.ru,nazarastron2002@mail.ru}

\altaffiltext{9}{Present address: 
		 Astronomy Department, 
		 University of Texas, 
		 Austin, TX 78712-0259;
                 gaskell@astro.as.utexas.edu}

\altaffiltext{10}{Present address: 
	          Astronomy Department, MSC 4500,
		  New Mexico State University, 
		  PO BOX 30001, La Cruces, NM 88003-8001}

\altaffiltext{11}{Department of Astronomy,
		 University of Michigan,
		 500 Church St., 
		 Ann Arbor, MI 48109-1040}

\altaffiltext{12}{Homer L. Dodge Department of Physics and Astronomy,
	    	  The University of Oklahoma,
  		  440 W. Brooks St.,
  		  Norman, OK 73019}
	
\altaffiltext{13}{Institute of Astronomy, 
		 School of Science, 
		 University of Tokyo,
	 	 2-21-1 Osawa, Mitaka, 
		 Tokyo 181-0015, Japan;
		 minezaki, yoshii@ioa.s.u-tokyo.ac.jp}

\altaffiltext{14}{Ohio University,
		  Department of Physics and Astronomy,
		  Athens, OH 45701-2979}

\altaffiltext{15}{Department of Astronomy, 
		  School of Science, 
		  University of Tokyo,
  		  7-3-1 Hongo, Bunkyo-ku, 
		  Tokyo 113-0013, Japan}

\altaffiltext{16}{Isaak Newton Institute of Chile,
	          Crimean Branch, Ukraine}

\altaffiltext{17}{Steward Observatory, 
		The University of Arizona, 
		933 North Cherry Avenue, 
         	Tucson, AZ 85721}

\altaffiltext{18}{Present address:
	         Department of Physics and Astronomy, 
		 Tufts University, 
		 Medford, MA 02155}

\altaffiltext{19}{Research Center for the Early Universe, 
		School of Science,
        	University of Tokyo, 
		7-3-1 Hongo, Bunkyo-ku, 
		Tokyo 113-0033, Japan}

\begin{abstract}

We present the first results from a high sampling rate, multi-month
reverberation mapping campaign undertaken primarily at MDM Observatory
with supporting observations from telescopes around the world.  The
primary goal of this campaign was to obtain either new or improved
\Hbeta\ reverberation lag measurements for several relatively low
luminosity AGNs.  We feature results for NGC~4051 here because, until
now, this object has been a significant outlier from AGN scaling
relationships, e.g., it was previously a $\sim$2--3$\sigma$ outlier on
the relationship between the broad-line region (BLR) radius and the
optical continuum luminosity --- the $R_{\rm BLR}$--$L$ relationship.
Our new measurements of the lag time between variations in the continuum
and \Hbeta\ emission line made from spectroscopic monitoring of NGC~4051
lead to a measured BLR radius of $R_{\rm BLR} = 1.87^{+0.54}_{-0.50}$
light days and black hole mass of $M_{\rm BH} = (1.73^{+0.55}_{-0.52})
\times 10^{6}M_{\odot}$.  This radius is consistent with that expected
from the $R_{\rm BLR}$--$L$ relationship, based on the present
luminosity of NGC~4051 and the most current calibration of the relation
by \citet{Bentz09b}.  We also present a preliminary look at
velocity-resolved \Hbeta\ light curves and time delay measurements,
although we are unable to reconstruct an unambiguous velocity-resolved
reverberation signal.
\end{abstract}



\section{INTRODUCTION}

Recent theoretical and observational studies have provided strong
evidence suggesting a connection between supermassive black hole (SMBH)
growth and galaxy evolution
\citep[e.g.,][]{Bennert08,Somerville08,Shankar09,Hopkins09}.  To better
understand this connection, we need more direct measurements of SMBH
masses across cosmological distances.  Unfortunately, measuring SMBH
masses directly with dynamical methods requires high angular resolution,
so use of these methods is limited to relatively nearby sources.  This
resolution problem can be obviated by studying SMBHs in type 1 active
galactic nuclei (AGNs).  In this case, the technique of reverberation
mapping \citep{Blandford82,Peterson93} may be applied to measure the
SMBH mass, as has been done for more than 3 dozen type 1 AGNs to date
\citep[e.g., see recent compilation by][]{Peterson04}.

Reverberation mapping relies on time resolution rather than angular
resolution, since it takes advantage of the presence of a time delay,
$\tau$, between continuum and emission line flux variations observed
through spectroscopic monitoring.  This time delay corresponds to the
light travel time across the broad-line region (BLR), and thus
measurements of $\tau$ provide an estimate of the size of the region,
$R_{\rm BLR} = c\tau$.  Because the BLR gas is well within the sphere of
influence of the black hole and studies have provided evidence for
virialized motions within this region \citep[e.g.,][and references
therein]{Peterson04}, $R_{\rm BLR}$ can be related to the mass of the
SMBH through the velocity dispersion of the BLR gas.

Although stellar and gas dynamical modeling and reverberation mapping
are invaluable for measuring SMBH masses directly, these methods are
observationally taxing, due to resolution requirements for dynamical
methods and time requirements for reverberation mapping campaigns.  It
is currently impossible to measure SMBH masses directly for large
statistical samples of galaxies.  However, these direct mass
measurements have led to the discovery of scaling relationships that
relate SMBH mass to other galaxy or AGN observables that can be used to
investigate the connection between SMBH mass and galaxy evolution.  In
particular, some relations show connections between properties of the
SMBH (i.e., its mass) and global properties of the host galaxy.
Examples include the correlation between SMBH mass and bulge/spheroid
stellar velocity dispersion, i.e. the $M_{\rm BH}$--$\sigma_{\star}$
relation for AGNs \citep{Gebhardt00b, Ferrarese01, Onken04, Nelson04}
and quiescent galaxies \citep{Ferrarese00, Gebhardt00a, Tremaine02}, and
the correlation between SMBH mass and galaxy bulge luminosity
\citep{Kormendy95,Magorrian98,Wandel02,Graham07,Bentz09a}.  Other
relations connect various AGN properties.  One such relation is the
correlation between black hole mass and optical luminosity
\citep{Kaspi00,Peterson04}, which relates directly to the accretion
rates of AGNs.  There is also a correlation between BLR radius and AGN
luminosity, i.e., the $R_{\rm BLR}$--$L$ relation
\citep{Kaspi00,Kaspi05,Bentz06a, Bentz09b}, which has proven to be very
powerful for making indirect SMBH mass estimations from single-epoch
spectra \citep[e.g.,][]{Vestergaard02,Vestergaard04,Corbett03,
Kollmeier06, Vestergaard08,JShen08,YShen08,Fine08}.  These indirect mass
estimates can then be related to other properties of the host galaxy
through direct measurements or separate scaling relations.

Although scaling relations have become widely used for statistical
studies, it is important to understand that the indirect mass estimates
determined by these relations are only as reliable as the direct mass
measurements used to calibrate them.  Therefore, establishing a secure
calibration across a wide dynamic range in parameter space and better
understanding any intrinsic scatter in these relations is essential.  To
accomplish this, we must continue to make new direct measurements as
well as to check previous results that are, for one reason or another,
suspect.

NGC 4051, an SAB(rs)bc galaxy with a narrow-line Seyfert 1 (NLS1)
nucleus at redshift $z=0.00234$, is a case in point.  Measurements of
the BLR radius and optical luminosity \citep{Peterson00b, Peterson04}
place it above the $R_{\rm BLR}$--$L$ relation, i.e., the BLR radius is
too large for its luminosity \citep[cf. Figure 2 of][]{Kaspi05}. It also
appears to be accreting mass at a lower Eddington rate than other NLS1s
\citep[cf. Figure 16 of][]{Peterson04}.  These two anomalies together
suggest that perhaps the BLR radius has been overestimated by
\citet{Peterson00b,Peterson04}; indeed an independent reverberation
measurement of the BLR radius in NGC 4051 by \citet[][hereafter
S03]{Shemmer03} is about half the value measured by \citet[][hereafter
P00]{Peterson00b}.  Furthermore, neither the P00 nor S03 data sets are
particularly well sampled on short time scales, so neither set is
suitable for detection of smaller time lags (e.g., $\lesssim$2--3 days).
In addition, \citet{Russell03} reports a Tully-Fisher distance to
NGC~4051 that is $\sim$50\% larger than that inferred from its redshift
(i.e., 15.2 Mpc versus 10.0 Mpc, respectively).  This suggests that the
luminosity derived in past studies from the redshift could be an
underestimate and might also be a contributing factor to the placement
of NGC~4051 above the $R_{\rm BLR}$--$L$ relation.

In this work, we present an analysis of new, optical spectroscopic and
photometric observations of NGC~4051, which represent the first results
from a densely sampled reverberation mapping campaign that began in
early 2007.  The campaign spanned more than 4 months, during which time
we consistently obtained multiple photometric observations per night and
spectroscopic observations nearly every night from a combination of five
different observatories around the globe.  The immediate goal of this
campaign is to remeasure the \Hbeta\ reverberation lag measurements for
several objects on the low-luminosity end of $R_{\rm BLR}$--$L$ scaling
relationship with poorly determined reverberation lags and,
consequently, poorly determined black hole masses.  We will also add to
the overall sample of reverberation mapped AGNs by measuring lags for
new objects.  Another goal for this extensive collection of homogeneous
data is to reconstruct the observed response of the \Hbeta\ emission
line to an outburst from the variable continuum source by modeling the
response as a function of both line-of-sight velocity and light travel
time, i.e., a ``velocity--delay map'' \citep[for a tutorial,
see][]{Peterson01,Horne04}.  Creation of a velocity--delay map will provide
novel insight into the structure and kinematics of the BLR.  Though we
have not yet attempted to reconstruct a full velocity--delay map, we
present preliminary velocity-resolved lag measurements for NGC~4051.
Complete results for NGC~4051 and other campaign targets will be
presented in future work.

\section{Observations and Data Analysis}

Most data acquisition and analysis practices employed here follow
closely those described by \citet{Denney06} and laid out by
\citet{Peterson04}.  The reader is referred to these works for
additional details and discussions.  Throughout this work, we assume the
following cosmological parameters: $\Omega_{m}=0.3$,
$\Omega_{\Lambda}=0.70$, and $H_0 = 70$ km sec$^{-1}$ Mpc$^{-1}$.

\subsection{Spectroscopy}

\begin{figure}
\figurenum{1}
\epsscale{1}
\plotone{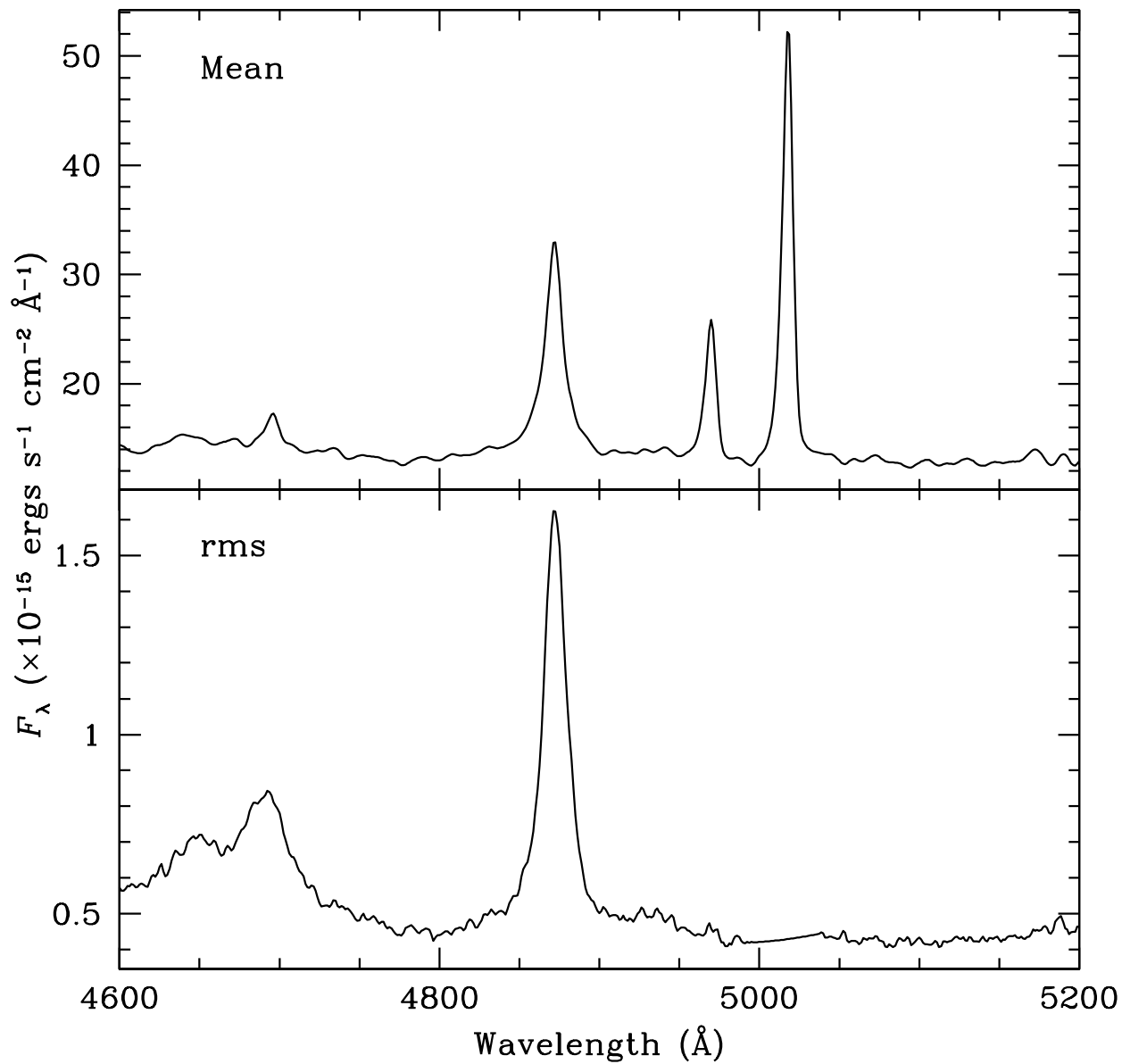}

\caption{Mean and rms spectra of NGC~4051 from MDM observations.  The
rms spectrum was formed after removing the [O\,{\sc iii}]\,$\lambda
4959$ and [O\,{\sc iii}]\,$\lambda 5007$ narrow emission lines. The
variability signature of \Hbeta\ is clearly visible in the rms spectrum,
and the large increase in rms flux shortward of 4800 \AA\ is due to
variations in the broad He\,{\sc ii}\,$\lambda 4686$ emission line.}

\label{fig:meanrms}
\end{figure}

Spectra of the nuclear region of NGC~4051 were obtained from both the
1.3-meter telescope at MDM Observatory and the 2.6-meter Shajn telescope
of the Crimean Astrophysical Observatory (CrAO).  The MDM observations
utilized the Boller and Chivens CCD spectrograph, where 86 observations
were taken over the course of 89 nights between JD2454184 and JD2454269,
targeting the H$\beta\,\lambda 4861$ and [O\,{\sc iii}]\,$\lambda\lambda
4959, 5007$ emission line region of the optical spectrum.  The position
angle was set to $0 \degr$, with a slit width of 5\farcs0 projected on
the sky, resulting in a spectral resolution of 7.6~\AA\ across this
emission-line region.  Figure \ref{fig:meanrms} shows the mean and rms
spectra of NGC~4051 based on the MDM observations.  We acquired 22 CrAO
spectra over 34 nights between JD2454266 and JD2454300 with the Nasmith
spectrograph and SPEC-10 $1340 \times 100$ pixel CCD.  For these
observations a 3\farcs0 slit was utilized, with a $90 \degr$ position
angle.  Spectral wavelength coverage for this data set was from
$\sim$3800--6000~\AA, with a dispersion of 1.8~\AA/pix and a spectral
resolution of 7.5~\AA.  Note that (1) the dispersion varies with
wavelength: the value 1.8 \AA/pix is given for 5100\AA, and (2) the real
wavelength coverage is slightly greater than given but the red and blue
edges of the CCD frame are unusable (too low S/N ratio) because of
vignetting.

A relative flux calibration of each set of spectra was performed based
on the constancy of the narrow [\oiii]\,$\lambda 5007$ line
flux. Because this line emission originates in the extended, low-density
narrow line region, it can be assumed constant over the timescale of
this campaign and therefore serves as the basis for a reliable relative
flux calibration.  However, the data quality is not identical from night
to night due to, e.g., seeing, weather conditions, atmospheric
transparency, etc.  This affects not only the integrated line flux in
each observation, but also properties of the spectrum such as S/N and
resolution.  Consequently, we employ the spectral scaling algorithm of
\citet{vanGroningen92} for the [\oiii]\,$\lambda 5007$ flux calibration.
This algorithm determines the best scaling through $\chi^{2}$
minimization of residuals rather than simply calculating a simple
multiplicative scale factor to scale the spectral fluxes.  Following
this method, we created a reference spectrum by averaging all spectra
from a given data set.  We then formed a difference spectrum by
subtracting the reference spectrum from each individual spectrum.  The
algorithm uses a least squares method to minimize the residuals of the
[\oiii]\,$\lambda 5007$ line flux in each difference spectrum by making
small zero-point wavelength calibration adjustments, correcting for
resolution differences, and applying a multiplicative scale factor to
the [\oiii]\,$\lambda 5007$ line flux of the individual spectrum.
Because this method is based on minimizing residuals between each
individual spectrum and the reference spectrum, there is a small
residual dispersion in the line fluxes after calibration.  This
dispersion is related to the data quality and the ability of the scaling
algorithm to mitigate night to night differences between individual
spectra, related largely to seeing and S/N.  Tests of the original
algorithm by \citet{vanGroningen92} estimated errors in the scaled
fluxes of better than 5\%, however, past studies employing somewhat
improved versions of this same scaling algorithm typically achieved
dispersions across a data set of $\sim$2\% \citep[e.g.,][]{Peterson95}.
We measure a dispersion of $\sim$1.5\%, demonstrating the high level of
homogeneity that we have been able to achieve in the current data set.

\subsection{Photometry}

In addition to spectral observations, we obtained supplemental $V$-band
photometry from the 2.0-m Multicolor Active Galactic NUclei Monitoring
(MAGNUM) telescope at the Haleakala Observatories in Hawaii, the 70-cm
telescope of the CrAO, and the 0.4-m telescope of the University of
Nebraska.

The MAGNUM observations were imaged with the multicolor imaging
photometer (MIP) as described by \citet{Kobayashi98a,Kobayashi98b},
\citet{Yoshii02}, and \citet{Yoshii03}.  Photometric fluxes measured
from 23 observations between JD2454182 and JD2454311 within an aperture
of 8\farcs3.  Photometric reduction of NGC~4051 was similar to that
described for other sources by \citet{Minezaki04} and
\citet{Suganuma06}, except the host-galaxy contribution to the flux
within the aperture was not subtracted and the filter color term was not
corrected because these photometric data were later scaled to the MDM
continuum light curve (as described below).  Also, minor corrections (of
order 0.01 mag or less) due to the seeing dependence of the host-galaxy
flux were ignored.

The 76 CrAO photometric observations were collected between J245D4180
and JD2454299 with the AP7p CCD mounted at the prime focus of the 70-cm
telescope ($f=282$ cm).  In this setup, the $512 \times 512$ pixels of
the CCD field covers a 15\farcm\ $\times$ 15\farcm\ field of view.
Photometric fluxes were measured within an aperture of 15\farcs0.  For
further details of the CrAO $V$-band observations and reduction, see the
similar analysis described by \citet{Sergeev05}.

The University of Nebraska observations were conducted by taking and
separately measuring a large number of one-minute images ($\sim$20) each
of 28 nights between JD2454195 and JD2454290.  Details of the observing
and reduction procedure are as described by \citet{Klimek04}.
Comparison star magnitudes were calibrated following
\citet{Doroshenko05a,Doroshenko05b} and \citet{Chonis08}.  To minimize
the effects of variations in the image quality, fluxes were measured
through an aperture of radius 8 arcseconds.  The errors given for each
night are the errors in the means.

\subsection{Light Curves}
\label{S:lightcurves}

\begin{figure}
\figurenum{2}
\epsscale{1}
\plotone{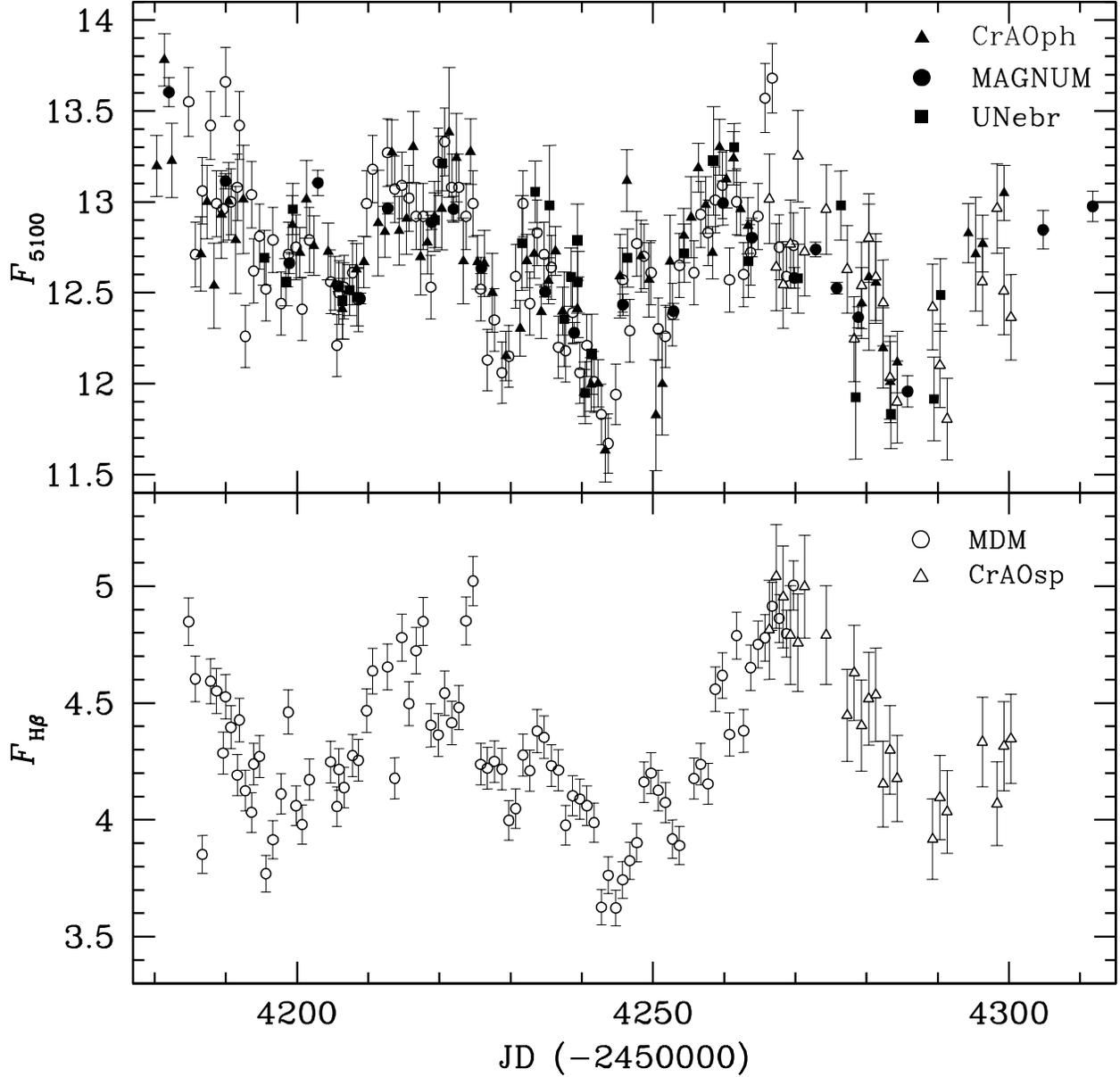}

\caption{Light curves showing complete set of observations from all four
sources.  The top panel shows the 5100 \AA\ continuum flux in units of
$10^{-15}$ erg s$^{-1}$ cm$^{-2}$ \AA$^{-1}$, while the bottom is the
H$\beta$ $\lambda$4861 line flux in units of $10^{-13}$ erg s$^{-1}$
cm$^{-2}$.  The open triangles (CrAOsp) correspond to spectroscopic
observations taken at CrAO, while the closed triangles (CrAOph)
represent photometric observations from CrAO.}

\label{fig:lcsep}
\end{figure}

Light curves of the \Hbeta\ flux were made based on integrated fluxes
measured in the MDM and CrAO spectra between 4815--4920~\AA\ and over a
linearly interpolated continuum defined between the average flux density
in each of the following regions blueward and redward of \Hbeta,
respectively: 4770--4780~\AA\ and 5090--5130~\AA.  The CrAO \Hbeta\
light curve was then scaled to the MDM light curve with a multiplicative
constant based on the average flux ratio between the four pairs of
closely spaced points in the MDM and CrAO \Hbeta\ light curves separated
by no more than 0.5 day.  This scaling is necessary to account for
differences in the amount of [\oiii]\,$\lambda 5007$ emission line flux
that enters the slit in the different data sets due to seeing and
aperture affects.  The lower panel of Figure \ref{fig:lcsep} shows the
\Hbeta\ light curve derived from both data sets after scaling the CrAO
fluxes to those measured from MDM spectra.

A continuum light curve was created with observations from each $V$-band
photometric data set and the average continuum flux density measured
over 5090--5130~\AA\ (i.e., rest frame $\sim$5100~\AA) in each spectrum
of the spectroscopic data sets.  First, the multiplicative scale factor
determined for above to scale the CrAO \Hbeta\ fluxes to the MDM light
curve was also applied to the CrAO continuum fluxes, since the
calibration of these fluxes is also susceptible to the same seeing and
aperture affects as the \Hbeta\ flux calibration.  Next, this light
curve as well as the individual photometric light curves (see
Fig. \ref{fig:lcsep}, upper panel) were scaled, one by one, to the same
flux scale as the MDM light curve by making an additive, relative flux
adjustment to each.  This additive correction is necessary for both
spectroscopic and photometric data because of differences in host galaxy
starlight that enters the different aperture sizes of the various data
sets.  For the photometric observations, there is an additional
component (also additive) due to the larger width of the filter
bandpass.  Each light curve was merged with the parent light curve to
which it was scaled before the next light curve was scaled, thus
building up a larger, more well-sampled light curve in the following
order: MDM, MAGNUM, CrAO photometry, University of Nebraska, and CrAO
spectroscopy. This was done so that the smaller, and in some cases
shorter, light curves could be scaled to a longer and more densely
sampled parent light curve.  The scale factor applied to each secondary
light curve to scale it to its parent light curve was calculated based
on the difference between a linear least squares fit to this light curve
and to the parent light curve before it, starting with the MDM light
curve as the initial parent light curve.  The fits to each light curve,
both parent and secondary were limited to using only observations within
the same overall temporal range, so that, when necessary, the beginning
and/or ends of the light curves were truncated during the fitting
process.

Assignment of uncertainties to the photometric fluxes is described above
in the text or in references describing the photometric data sets.
However, we calculated the uncertainties in the MDM and CrAO
spectroscopic fluxes after creating the light curves for these data sets
but before intercalibration.  Typically, we determine uncertainties in
our light curve flux measurements by applying a mean fractional error to
all points. This fractional error is determined by comparing the average
flux difference between closely spaced pairs of observations, assuming
that flux differences across these short times scales are due to noise
rather than genuine variability.  Because real variability has been
confirmed by \citet{Klimek04} to occur in NGC~4051 on time scales
shorter than 2 days, and our sampling rate is $\sim$1 day, on average,
we could not use this method to determine the relative errors on our
spectroscopic flux measurements for this object.  Instead, we took
advantage of the observations of other higher-luminosity AGNs that we
monitored as part of this larger campaign (i.e., same telescopes,
instrumental setup, and observing conditions; results in preparation).
Unlike NGC~4051, these objects neither exhibit variability on such short
time scales nor have such short measured lags.  Therefore, the
uncertainties assigned to the NGC~4051 spectroscopic observations seen
in Figure \ref{fig:lcsep} are an average of the uncertainties in the
flux measurements calculated as described above from closely spaced
observations (separations of $\lesssim$2.0 days) of these other objects
(e.g., typical fractional errors in the range $\sim0.12-0.21$ and
$\sim0.14-0.27$ for the continuum and line fluxes, respectively).

\begin{figure}
\figurenum{3}
\epsscale{1}
\plotone{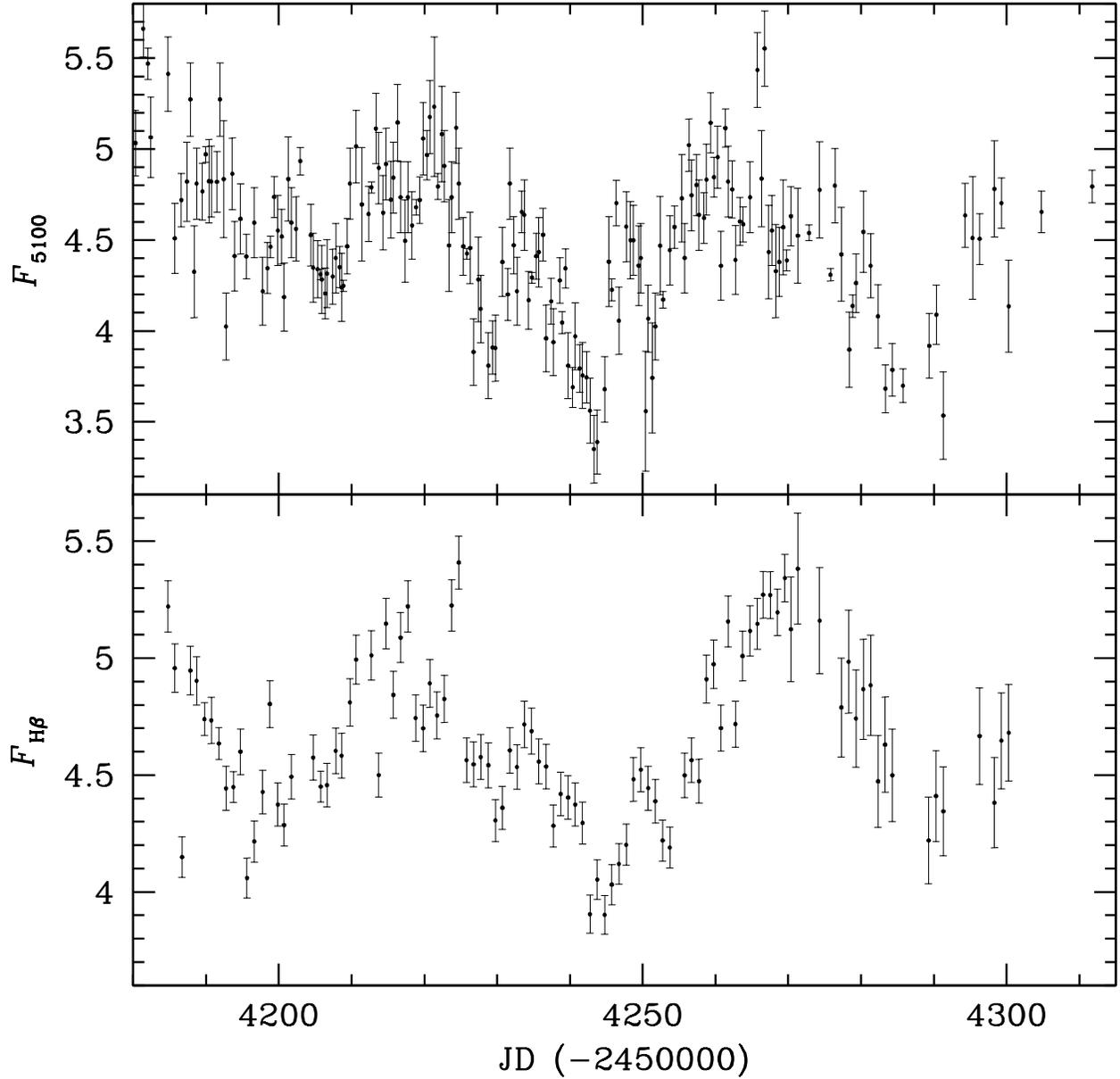}

\caption{Same as Fig. \ref{fig:lcsep} except data from all sources have
been merged and closely spaced observations binned such that weighted
averages were calculated for continuum observations separated by less
than 0.25 day and \Hbeta\ observations separated by less than 0.5 day.}

\label{fig:lc4cc}
\end{figure}

The merged continuum and \Hbeta\ light curves shown in Figure
\ref{fig:lc4cc} are used for the subsequent time series analysis.  These
differ from simply combining the individual light curves, shown in
Figure \ref{fig:lcsep}, in the following ways:
\begin{itemize}

\item First, we applied an absolute flux calibration to both light
curves by applying a single multiplicative scale factor determined from
the ratio of the [\oiii]\,$\lambda 5007$ emission line flux determined
by P00 to that measured in the reference spectrum used above for the
relative flux calibration.  Unlike the emission line flux in our
reference spectrum, the P00 flux measurement of $F($[\oiii]\,$\lambda
5007) = (3.91 \pm 0.12) \times 10^{-13}$ erg s$^{-1}$ cm$^{-2}$ was
taken from observations obtained under photometric conditions and,
consequently, $\sim$8\% larger than our measured value.  Additionally,
the P00 measurement was made employing observing strategies and
measurement practices similar to what we present here, thus validating
this direct comparison.  This additional flux calibration does not
affect the reverberation results but is necessary for accurately
measuring the 5100\AA\ continuum luminosity.

\item Second, we subtracted the host galaxy starlight contribution to the
continuum flux, determined using the methods of \citet{Bentz09b} to be
$F_{gal}$(5100\AA)$=(9.18 \pm 0.85) \times 10^{-15}$ erg s$^{-1}$
cm$^{-2}$ \AA$^{-1}$.

\item Third, we binned closely spaced observations as a weighted
average and applied this to continuum flux measurements separated by
$\leq$0.25 days and \Hbeta\ flux measurements separated by $\leq$0.5
days.

\end{itemize}

Fluxes for individual observations (i.e., before time binning) from all
sources are listed in Table \ref{tab:fluxes}.  Values listed represent
the flux of each observation after completing all flux calibrations
described above (i.e., relative calibration to intercalibrate all data
sets onto the MDM flux scale, followed by absolute calibration based on
the P00 [\oiii]\,$\lambda 5007$ line flux and removal of host starlight
contamination).  Column 1 gives the Julian Date of each observation.
The 5100~\AA\ continuum or $V$-band flux and integrated \Hbeta\ flux are
given in columns (2) and (3), respectively, and column (4) lists the
source of each measurement.  Photometric and spectroscopic observations
from CrAO can be differentiated by noting that no \Hbeta\ flux values
are present for photometric observations.

Table \ref{tab:lcstats} displays statistical parameters describing the
final light curves shown in Figure \ref{fig:lc4cc}.  Column (1) gives
the spectral feature represented by each light curve, and the number of
data points in each light curve is shown in column (2).  Columns (3) and
(4) are mean and median sampling intervals, respectively, between data
points. The mean flux with standard deviation is given in column (5),
while column (6) shows the mean fractional error in these fluxes.
Column (7) gives the excess variance, calculated as

\begin{equation}
F_{\rm var} = \frac{\sqrt{\sigma^2 - \delta^2}}{\langle f \rangle}
\end{equation}

\noindent where $\sigma^2$ is the variance of the observed fluxes,
$\delta^2$ is their mean square uncertainty, and $\langle f \rangle$ is
the mean of the observed fluxes \citep{Rodriguezpascual97, Edelson02}.
Finally, column (8) is the ratio of the maximum to minimum flux in the
light curves.

\subsection{Time Series Analysis}
\label{S:lagresults}

A times series analysis of the continuum and \Hbeta\ light curves was
performed to determine the mean light travel time lag between continuum
and emission line variations.  We used two cross correlation schemes
designed for data sets with uneven time sampling:
\begin{enumerate}

\item An interpolation scheme \citep{Gaskell86,Gaskell87,White94} with an
interval of 0.2 day.  A cross-correlation function (CCF) is constructed
from the mean value of the correlation coefficient, $r$, computed from
cross correlating both the interpolated line light curve with the
original continuum light curve and then the interpolated continuum light
curve with the original line light curve, a process during which a range
of possible lags, $\tau$, are imposed on the \Hbeta\ light curve.

\item A time binning scheme \citep{Edelson88,White94} with a bin size of
  1.0 day.  Here, a discrete correlation function (DCF) is produced,
  which determines $r$ as a function of lag, similar to the CCF.  In
  this scheme, however, only the actual data are cross correlated, and
  the resulting values of $r$ for all discretely correlated pairs are
  binned as a function of lag.  The DCF that results gives the mean
  value of $r$ in each bin, where the corresponding uncertainty is
  assigned in a statistical manner \citep[see][]{White94}.  This method
  prevents possible spurious lag determinations that could potentially
  arise in the interpolation method due to under-sampling or large gaps
  in the data.
\end{enumerate}

\begin{figure}
\figurenum{4}
\epsscale{1}
\plotone{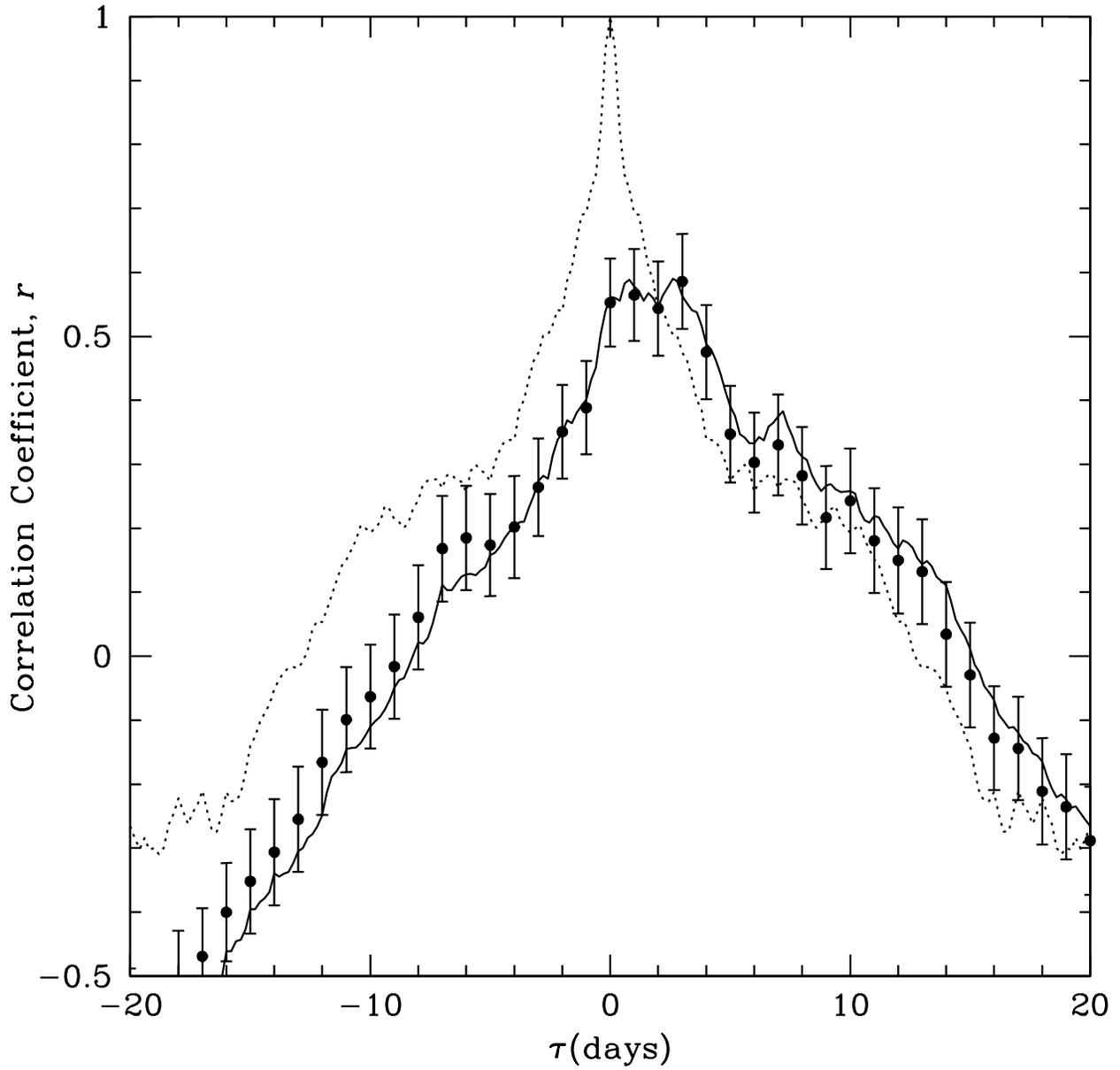}

\caption{Cross correlation function (CCF; solid line), discrete
correlation function (DCF; filled circles), and auto correlation
function (ACF; dotted line) from time series analysis of the continuum
and H$\beta$ light curves of NGC~4051 shown in Fig. \ref{fig:lc4cc}.}

\label{fig:ccf}
\end{figure}

The resulting CCF and DCF are shown in Figure \ref{fig:ccf}, along with
the auto-correlation function (ACF) computed by cross correlating the
continuum with itself.  We characterize the time delay between the
continuum and emission line variations using two parameters derived from
the CCF; $\tau_{\rm peak}$ is the lag that corresponds to the largest
correlation coefficient, $r_{\rm max}$, and $\tau_{\rm cent}$ is the
centroid of the CCF based on all points with $r\geq 0.8r_{\rm max}$.
Time dilation corrected values of $\tau_{\rm peak}$ and $\tau_{\rm
cent}$ determined from the CCF in Figure \ref{fig:ccf} are given in
Table \ref{tab:results}.  Uncertainties in both lag parameters are
computed via model-independent Monte-Carlo simulations that employ the
bootstrap method of \citet{Peterson98}, with the additional
modifications of \citet{Peterson04}.

\section{Comparison with Previous Results}
\label{sec:compare}

Our measured \Hbeta\ lag of $\tau_{cent} = 1.87^{+0.54}_{-0.50}$ days
from this work is consistent, within the errors, to the most recent
results for this object by S03, who measured a lag of $\tau_{cent} = 3.1
\pm 1.6$ days.  It is not clear how meaningful a direct comparison might
be, however, because S03 measured the time delay between variations in
the $\sim$6800\AA\ continuum flux density and the integrated H$\alpha$
flux rather than between the $\sim$5100\AA\ continuum and \Hbeta.  We
also note that the median sampling rate of S03 was larger than our
measured lag, suggesting to us that the S03 light curves are
undersampled.  Furthermore, S03 only perform a cross correlation
analysis based on the DCF method, which sacrifices time resolution.

Our new time delay measurements are inconsistent, however, with the
previous measurement of $\tau_{cent} = 5.8^{+2.6}_{-1.8}$ days by
\citet{Peterson04} using data from P00.  These differences are unlikely
a luminosity effect, since the average luminosity states of NGC~4051
were similar during this and the P00 campaigns (log$\lambda L_{5100}$ =
41.82 and log$\lambda L_{5100}$ = 41.87, respectively)\footnote{The
average observed flux of the S03 campaign was within $\sim 10 \%$ that
of the Peterson et al. campaign as well.}.  Therefore, we carefully
re-examined the light curves used by P00 to better understand possible
causes for the observed inconsistency.

\citet{Netzer90b} suggested that the cause for a similar inconsistency
between lag measurements from two reverberation mapping campaigns of
NGC~5548 \citep{Netzer90a,Peterson91} was due to different continuum
variability timescales observed in the separate campaigns: longer
continuum variability timescales lead to larger lag measurements.
However, this explanation is unlikely to be the cause for the current
inconsistency between our measured lag and that of P00 because the
prominent variability timescales observed in both the P00 and current
continuum light curves are similar ($\sim$40--50 days).  Instead, the
simplest explanation for the inconsistency between our measured lag and
that of P00 is random error.  We investigated this possibility by
performing Monte Carlo simulations using the ``Subset 1'' \Hbeta\ and
5100\AA\ continuum flux light curves from P00 with the goal of
estimating the likelihood that a lag of $\tau_{cent}$ = 5.8 days would
be measured, even if the actual BLR radius of the \Hbeta\ emission was
2.7 light days, as expected from the \citet{Bentz09b} $R$--$L$ relation
for the average luminosity of NGC~4051 during this time period.  In each
simulation we created a simulated \Hbeta\ light curve by convolving a
modified continuum light curve with a transfer function that assumed a
BLR with a thin spherical shell geometry of radius 2.7 light days.  The
sampling was increased in the modified continuum light curve over that
of the original Subset~1 continuum light curve by interpolating between
the actual points on a 0.5 day scale.  Noise was added to the flux of
each interpolated point using a random Gaussian deviate. The size of
each deviate was based on the average uncertainty in flux of the closest
`real' continuum point on each side of the interpolated point.  The
simulated emission-line light curve was then sampled identically to the
original Subset 1 \Hbeta\ light curve.  We cross correlated this new
emission-line light curve with the original Subset 1 continuum light
curve to determine a reverberation lag.  The simulation was repeated
10,000 times, and lags were measured similarly for each iteration.  We
found that the average lag recovered was 2.7 days (reassuring, since
this was our input radius).  However, we were unable to reproduce even a
single lag of 5.8 days.  In fact, the largest lag our simulations
recovered was 4.0 days.  A couple of possibilites suggest themselves:
\begin{itemize}
\item The BLR has physically changed in the 11-year interval between the
time of P00's Subset 1 and the time of our recent campaign.  This is a
physical possibility since the dynamical timescale of the BLR in
NGC~4051 is $\ltsim 5$ years.

\item The P00 data are undersampled and there are really unresolved
variations occurring on timescales shorter than the typical sampling
interval of 2.2 days in Subset 1, the best-sampled part of the P00 light
curve.
\end{itemize}

We have no particular reason to believe the former possibility. However,
the latter is suggested by how P00 established the relative uncertainties
of their fluxes, namely by assuming that there are no true variations on
time scales shorter that the typical sampling time scales and that any
differences between closely spaced measurements reflect random errors
only, not true variability.  The estimates of the relative flux errors
in the P00 Subset 1 based on comparing measurements separated by 2 days
or less are about 3.2\% for both the contiuum and the line.  In our new
data set, obtained with the same instrument, we find relative errors of
about 1.4\% and 2.1\% in the continuum and the line, respectively, using
the method described in Section \ref{S:lightcurves}.  We conclude that the
flux uncertainties of the P00 data were overestimated due to short
timescale variability.

\begin{figure}
\figurenum{5}
\epsscale{1}
\plotone{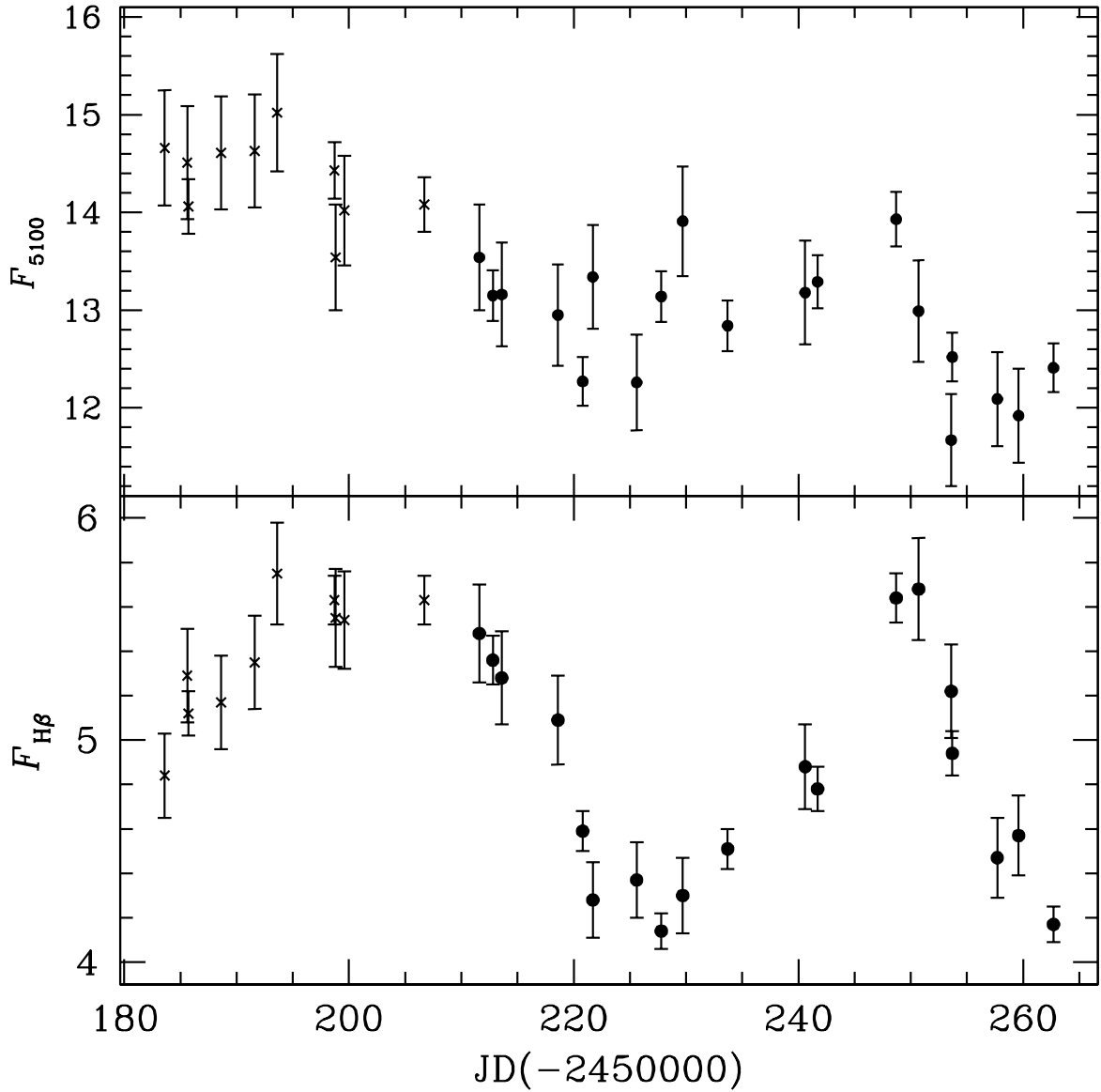}

\caption{Optical continuum and \Hbeta\ light curves reproduced from P00.
Points with X's represent those that were excluded from the new time
series analysis of this light curve described in Section
\ref{sec:compare}.  Units are the same as in Fig. \ref{fig:lcsep}.}

\label{fig:P00lc}
\end{figure}

Proceeding with the assumption that the P00 light curves are
undersampled, we isolated the portion of the light curves that has the
highest sampling across the sharpest features.  We made this selection
in an attempt to avoid occurrences of undersampling more complex
variability.  From the initial light curve, reproduced in Figure
\ref{fig:P00lc}, we removed the first 10 observations that exhibit a
broad inflection in the flux with a poorly defined peak.  We perform a
cross correlation analysis on these shortened light curves and determine
a shorter lag, $\tau_{cent}=3.5^{+3.7}_{-1.9}$ days.  This lag
determination is consistent with both the expected radius of 2.7 light
days from the $R$--$L$ relation, the current results, and the results of
S03.

If we continue with the assumption that the light curves of P00 are
undersampled, and the P00 flux uncertainties are overestimated, their
assigned uncertainties would act to decrease the significance of short
timescale variability, likely attributing it to noise instead.  If we
impose the average continuum flux uncertainty measured from our current
data set (given above) on the shortened P00 continuum light curve, we
can further improve the precision of this revised lag measurement to
$\tau_{cent}=3.5^{+3.2}_{-1.5}$ days.

We then conducted another simulation in which we applied the sampling
rate from the P00 light curves to the light curves from this work.  By
undersampling our current light curves, we can determine the probability
that undersampling could lead to an overestimated lag similar to that
measured by P00.  This type of simulation can provide further evidence
that the lag measured by P00 was an overestimate and a consequence of
undersampling.  At the same time, it could diminish the possibility that
the discrepancy in lag measurements is due to a difference in the
physical conditions or structure of the BLR during the P00 campaign
compared to the present.  Using the continuum and \Hbeta\ light curves
shown in Figure \ref{fig:lc4cc} as the starting point, we modified them
similarly to the continuum light curve described for our first set of
simulations (i.e., increasing the sampling by interpolating between data
points and adding noise to these points with a Gaussian deviate), but
this time we interpolated both the continuum and emission line light
curves from this work on a 0.1 day interval.  We then drew sample light
curves from this parent light curve with the same length and sampling
pattern as the full P00 light curve shown in Figure \ref{fig:P00lc}.  We
applied the same cross-correlation analysis (as described in Section
\ref{S:lagresults}) to measure lags from these sample light curves.  The
parent light curves cover a longer time span than the sample light
curves and therefore allow for multiple iterations of sample light
curves to be chosen from different subsets of the parent light curves.
The first iteration of sample light curves are created from the subset
of the parent light curves where the beginning points match up, but the
ends of the parent light curves are discarded.  We then build up
multiple iterations by shifting the starting point of the sample light
curve in time by one time step, i.e., 0.1 day, across the parent light
curves.  In this way, we were able to build up 330 sample light curves,
where in the last iteration, the sample light curves begin in the middle
of the parent light curves, but both sets of light curves end at the
same time.  Based on the cross correlation analysis from these 330
sample light curves, the probability of measuring $\tau_{\rm cent}
\geq$5.0 days is 0.6\% (2 out of 330), and the probability of measuring
$\tau_{\rm cent} \geq$3.5 days (i.e, the lag we calculated above from
only a portion of the P00 light curve) is $\sim$8\% (25 out of 330).  We
conclude that undersampling is at least a plausible explanation for the
difference between the P00 results and those reported here.

\section{Black Hole Mass}

Applying virial assumptions to the reverberating gas in the BLR, the
mass of the black hole can be defined by

\begin{equation}
M_{\rm BH} = \frac{f c \tau (\Delta V)^2}{G},
\end{equation}

\noindent where $\tau$ is the measured emission-line time delay, so that
c$\tau$ represents the BLR radius, and $\Delta V$ is the BLR velocity
dispersion \citep{Peterson04}.  The dimensionless factor $f$ depends on
the structure, kinematics, and inclination of the BLR and is of order
unity.

We estimate the BLR velocity dispersion from the line width of \Hbeta\
emission line.  The line width can be characterized by either the FWHM
or the line dispersion, i.e., the second moment of the line profile.
The FWHM and the line dispersion, $\sigma_{line}$, were measured from
both the mean and the rms spectra of NGC~4051 shown in Figure
\ref{fig:meanrms}.  Here, we have measured both quantities and their
uncertainties employing methods described in detail by
\citet{Peterson04}.  All measured values of the \Hbeta\ line width are
listed in Table \ref{tab:results}.  Typically, the narrow-line emission
component of the line should be removed before measuring the line width
\citep[see][]{Denney08}; however, this component could not be reliably
isolated from the rest of the line profile in this object.  As a result
the line widths measured in the mean spectrum, particularly the FWHM,
are less reliable for these purposes than the widths measured from the
rms spectrum\footnote{Since only BLR emission varies in response to the
ionizing continuum on reverberation timescales, flux contributions from
the narrow-line component will not contaminate the line width
measurement in the rms spectrum.}.
	
We calculate the black hole mass for NGC~4051 using $\tau_{\rm cent}$,
for the time delay, $\tau$, and the line dispersion, $\sigma_{\rm
line}$, measured from the H$\beta$ emission line in the rms spectrum,
for the emission-line width, $\Delta V$.  We utilize the calibration of
the reverberation mass scale of \citet{Onken04} for this choice of lag
and line width parameters and therefore adopt a scale factor value of
$f$=5.5.  We then use equation 2 to estimate the black hole mass of
NGC~4051 to be $M_{\rm BH} = (1.73^{+0.55}_{-0.52})\times
10^{6}M_{\odot}$.  Here, statistical and observational uncertainties
have been included, but intrinsic uncertainties from sources such as
unknown BLR inclination cannot be accurately ascertained.

\citet{Marconi08} have considered the effect of radiation pressure on
SMBH mass estimates and provide a new prescription for calculating the
SMBH mass that includes a correction factor to account for radiation
pressure.  Radiation pressure acts to partially counteract the force of
gravity on the BLR gas, since the outward radiation force has the same
radial dependence (${\it r}^{-2}$) as the inward gravitational force.
As a result the BLR gas motions are effectively under the influence of
an apparently smaller SMBH mass, which leads to an underestimate of the
true SMBH mass.  \citet{Marconi08} determine that although taking
radiation pressure into account is most important for black holes
radiating near the Eddington limit, it should be considered even for
systems with $L < L_{\rm Edd}$.  On the other hand, in a comparison
study of Type 1 versus Type 2 black hole masses determined with
independent methods, \citet{Netzer09} sees better agreement between the
mass and $L/L_{\rm Edd}$ distributions of these two populations if
radiation pressure forces are neglected.  Netzer concludes that either
the effects of radiation pressure on the BLR gas in these objects is
negligible or that BLR column densities must be significantly larger,
i.e., $N_{\rm H}\gtrsim 10^{24} \rm cm^{-2}$, than assumed by Marconi et
al. (2008; $N_{\rm H}=10^{23} \rm cm^{-2}$).  In a more recent paper,
\citet{Marconi09} reinvestigate the results of \citet{Netzer09} and
support their findings on the dependence of the effect of radiation
pressure on column density but conclude that, until it is possible to
determine the nature of the apparent dependence of $N_{\rm H}$ on source
properties (e.g., $L/L_{\rm Edd}$), one should always consider the
possibility that radiation forces are important, and black hole masses
should consequently be determined using the correction of
\citet{Marconi08}.

The importance of radiation pressure forces on black hole mass
determinations is still under debate.  Therefore, in addition to our
virial mass estimate for NGC~4051 given above, we also estimate the SMBH
mass in NGC~4051 taking radiation pressure into consideration
\citep[cf. equation 6 of][]{Marconi08}, with $f=3.1 \pm 1.4$ and
log\,$g=7.6 \pm 0.3$, which are derived by Marconi et al. from fits to
SMBH masses from reverberation mapping studies.  With these scale
factors and the same line width and BLR radius measurements used above,
we calculate a mass of $M_{\rm BH-rad} = (1.24^{+0.57}_{-0.56}) \times
10^{6}M_{\odot}$.  Contrary to the expectations from the physical
arguments, we calculate a mass smaller than that determined in the case
where we did not consider the effect of radiation pressure.  Because
NGC~4051 is a low-luminosity AGN radiating well below the Eddington
limit ($L/L_{\rm Edd} = 0.030$), the correction to the mass due to
radiation pressure is small (smaller even than the formal observational
uncertainties on the mass), adding only $0.26 \times 10^6 M_{\odot}$ to
the mass.  However, this radiation-corrected mass estimate is smaller
than our original estimate because the value of $f$ derived by Marconi
et al. is a factor of 1.8 smaller than the \citet{Onken04} value we
adopted above.  Because this scale factor was derived in a statistical
sense by both Onken et al. and Marconi et al., the difference in $f$
values between these two studies has the potential to affect the mass of
an individual object more than would be expected for a statistical
sample, particularly for the low accretion-rate objects that need only a
small radiation pressure correction.

\section{Velocity-Resolved Investigation}
\label{S:velresInvest}

The lag measurements between the continuum and \Hbeta\ emission in Table
\ref{tab:results} represent the average time delay across the BLR.
Because the BLR is an extended region and the velocity of the gas is
most likely a function of position, gas in different locations of the
BLR should respond to variations in the ionizing continuum flux on
slightly different time scales.  The observable result should be a
difference in the reverberation lag measurement in different parts of
the line profile (i.e., separated in velocity space).  Velocity-resolved
reverberation mapping thus gives us information about the kinematics of
gas in the BLR.  Previous studies of time delay differences between
multiple emission lines and the velocity dependence of the lag within a
single emission line have shown that the BLR is virialized and commonly
contains an additional inflow component \citep[e.g.,][]{Gaskell88,
Koratkar91c, Korista95, Done96, Welsh07, Bentz08}; however, the creation
of full velocity--delay maps \citep[see][]{Horne04} is an aspect of the
reverberation mapping technique that has not yet been fully realized
(though see \citealt{Horne91}, \citealt{Done96}, \citealt{Ulrich96}, and
\citealt{Kollatschny03} for previous attempts).  By resolving the
velocity-dependent reverberation response of the BLR better than has
been done in the past, we can reconstruct and analyze the
velocity--delay map to gain further insights into the geometry and
kinematics of the BLR.

\begin{figure}
\figurenum{6}
\epsscale{1}
\plotone{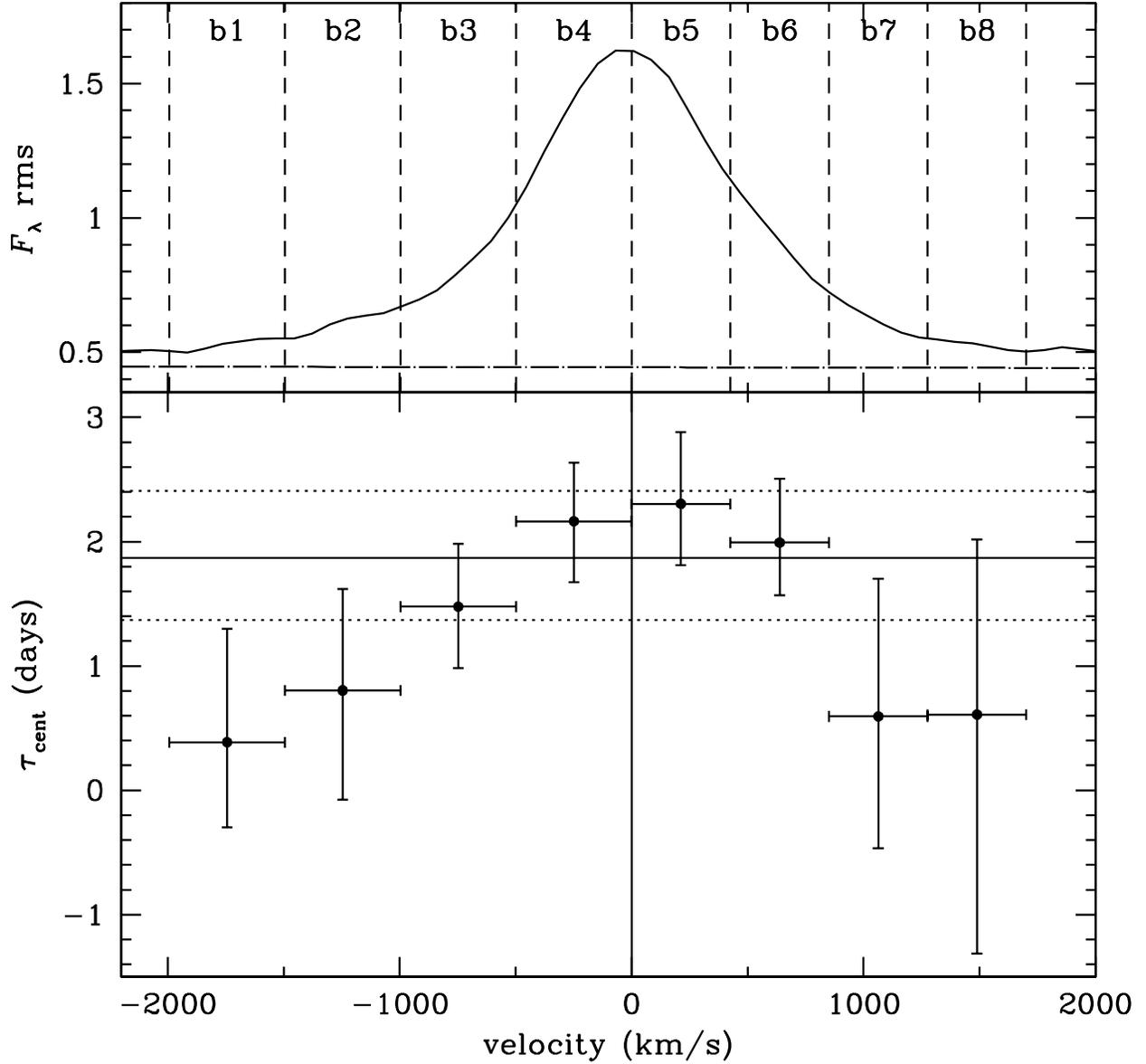}

\caption{Velocity-resolved \Hbeta\ rms spectrum profile (top) and
time-delay measurements (bottom) for NGC~4051.  Vertical dashed lines
plotted on the line profile (top) and error bars on the lag measurements
in the velocity direction (bottom) show the bin size, with each bin
labeled by number in the top panel.  Error bars on the lag measurements
are determined similarly to those for the mean BLR lag.  The horizontal
solid and dotted lines in the bottom panel show the mean BLR centroid
lag and associated errors, calculated in Section \ref{S:lagresults},
while the horizontal dotted-dashed line in the top panel represents the
linearly-fit continuum level.  Flux units are the same as in
Fig. \ref{fig:meanrms}.}

\label{fig:reslags}
\end{figure}

We searched for a velocity-dependent reverberation signal by dividing
the \Hbeta\ emission line flux into 8 velocity-space bins.  The blue and
red sides of the line were separately divided into 4 bins of equal
velocity width, covering only the most variable portions of the line
profile, i.e., the outer-most wavelength boundaries were reduced from
those considered for the analysis of the full profile to only include
flux within the range 4840--4900\AA\ (roughly $\pm$2,000 km s$^{-1}$).
Light curves were created from measurements of the integrated \Hbeta\
flux in each bin and then cross correlated with the continuum light
curve following the same procedures described above.  The top panel of
Figure \ref{fig:reslags} shows the division of the \Hbeta\ line profile
from the rms spectrum into the eight velocity bins, and the bottom panel
shows the lag measurements for each of these bins.  Error bars in the
velocity direction represent the bin width.  The evidence for a
velocity-stratified BLR response to continuum variations is present, but
marginal.  In particular, the lags measured for bins 1-2 and 7-8 are
consistent with zero and might simply reflect correlated errors due to
continuum contamination.  Although the shape of the velocity-resolved
signal in Figure \ref{fig:reslags} supports our virial assumptions,
since the higher velocity gas varies on shorter time scales than the low
velocity gas, there are no strong indications for either outflow or
inflow.  Outflow could be suggested by the larger lag measured in bin 5
(red side of the line) compared to bin 4 (blue side of the line), but
the difference is very marginal.  Even with the high sampling rate and
spectral resolution we achieved during this campaign, observing a
velocity-resolved signal for NGC~4051 as clearly as that detected for
Arp~151 by \citet{Bentz08} would have been rather surprising for the
following reasons.  First, the precision with which we can measure a lag
is somewhat dependent on the median observational sampling interval,
which, for NGC~4051, was still not much shorter than the measured lag.
This indicates that in order to better resolve a velocity-dependent
signal, we need even higher time resolution for this object.  Second,
because the \Hbeta\ line is particularly narrow in this object, there
are only a few velocity resolution elements across the line.

\section{Discussion}

\begin{figure}
\figurenum{7}
\epsscale{1}
\plotone{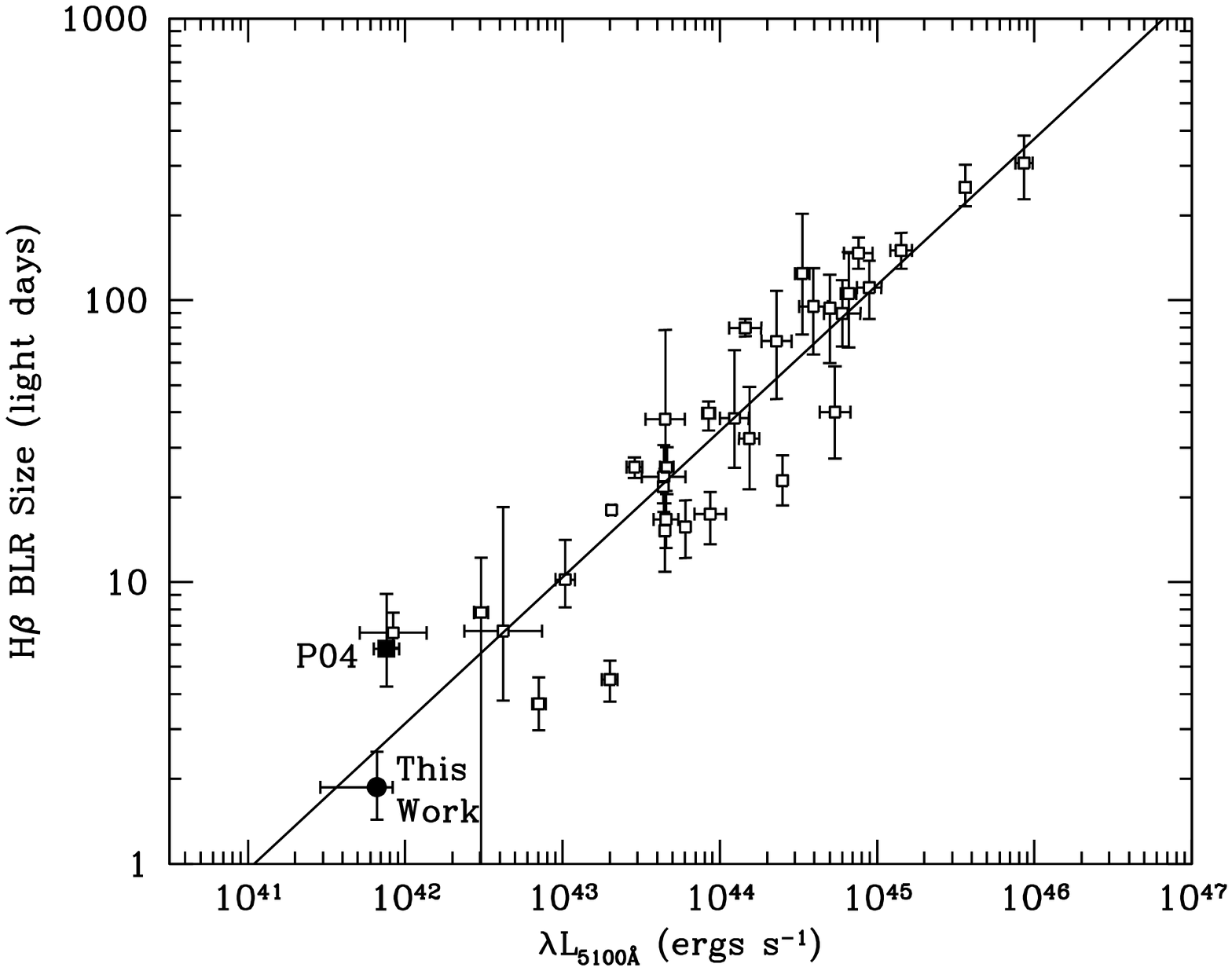}

\caption{Most recently calibrated $R_{\rm BLR}$--$L$ relation
\citep[][solid line]{Bentz09b}. The filled square shows the location of
NGC~4051 based on results from \citet{Peterson04} and used by Bentz et
al.  The filled circle shows the new lag measurement of 1.87 days
presented in this work at the luminosity calculated using the
Tully-Fisher distance to NGC~4051 of \citet{Russell03}: the error bar in
luminosity reflects both the range of flux variations, as for all of the
other data points, plus the uncertainty due to the distance, added in
quadrature.  The error bar is asymmetric, as we favor the larger
distance.  Open squares represent other objects from \citet{Bentz09b}.}

\label{fig:B08rlrelation} 
\end{figure}

Based on our simulations, reanalysis of the light curves from P00, and
the additional arguments we presented above, we conclude that the
inconsistency between the past measurements of the \Hbeta\ reverberation
lag in NGC~4051 and our present measurements most likely results from an
overestimation of the lag by P00.  Therefore, we adopt the results from
the current reverberation campaign over previous campaign results
measuring this lag in NGC~4051.  Using the lag measurement of
$\tau_{cent} = 1.87^{+0.54}_{-0.50}$ days presented here and the
Tully-Fisher distance of \citet{Russell03}, NGC~4051 is no longer an
outlier on the $R_{\rm BLR}$--$L$ relationship.  Figure
\ref{fig:B08rlrelation} replicates the most recent version of this
relationship by \citet{Bentz09b} with both the previous and current lag
values of NGC~4051 marked.  Secure placement of low-luminosity objects
such as NGC~4051 on the $R_{\rm BLR}$--$L$ relationship is important for
supporting the extrapolation of this relationship to the even
lower-luminosity regime potentially populated by intermediate-mass black
holes.  Additional results from the present campaign (Denney et al., in
preparation), as well as results from a recent monitoring campaign at
the Lick Observatory \citep[e.g.,][]{Bentz08}, aim to further populate
this low-luminosity end of the $R_{\rm BLR}$--$L$ relationship, thus
solidifying the calibration in a relatively under-sampled region of the
relation.  A reliable calibration of this relationship is imperative for
large studies of black hole masses and galaxy evolution, since it allows
for the calculation of black hole masses from single-epoch spectra and
provides luminosity and radius estimates that help constrain parameter
space in the search for intermediate-mass black holes.

Our new results provide a measure of the BLR radius of NGC~4051 that is
closer to the value naively expected from its luminosity.  However,
these new results do not resolve the unexpected location of this object
on the $M_{\rm BH}$--$L$ relation: NLS1s tend to lie on a locus of this
relation with relatively high Eddington ratios \citep[$L/L_{\rm Edd}
\gtrsim 0.1$; see Figure 16 of][]{Peterson04}.  However, based on the
black hole mass of $M_{\rm BH} = (1.73^{+0.55}_{-0.52}) \times
10^{6}M_{\odot}$ that we have calculated, the Eddington ratio of
NGC~4051 is still only $L/L_{\rm Edd} = 0.030$.  It seems unlikely that
we have overestimated the mass of the black hole (and thus
underestimated $L/L_{\rm Edd}$), since our mass measurement is already a
factor of a few lower than predicted by the $M_{\rm
BH}$--$\sigma_{\star}$ relationship \citep{Nelson95, Ferrarese01}.  In
this case, the narrowness of the Balmer lines in the spectrum of
NGC~4051 might be due at least in part to the inclination of the BLR
relative to our line of sight --- indeed, inclination has been invoked
as one possible way to explain the NLS1 phenomenon since the early days
of research on these sources
\citep[e.g.,][]{Osterbrock85,Boller96}. From observations of the
narrow-line region in NGC~4051, \citet{Christopoulou97} estimate that
the inclination of this source is $\sim 50\deg$, near the maximum
expected for a Type 1 active nucleus in unified models.  It is entirely
reasonable to suppose that the BLR and accretion disk are at the same
inclination as the narrow-line region.  Even so, accounting for this
high inclination would increase the line width by only about a factor of
2, still within the NLS1 regime. Clearly further investigation is
required to understand the low value of $L/L_{\rm Edd}$ in NGC~4051
compared to other NLS1s.

\acknowledgements We would like to thank the anonymous referee for
useful suggestions that improved the clarity of this manuscript.  We
acknowledge support for this work by the National Science Foundation
though grant AST-0604066 to The Ohio State University.  CMG is grateful
for support by the National Science Foundation through grants AST
03-07912 and AST 08-03883.  MV acknowledges financial support from HST
grants HST-GO-10417, HST-AR-10691, and HST-GO-10833 awarded by the Space
Telescope Science Institute, which is operated by the Association of
Universities for Research in Astronomy, Inc., for NASA, under contract
NAS5-26555.  VTD acknoledges the support of the Russian Foundation for
Basic Research (project no. 06-02-16843) to the Crimean Laboratory of
the Sternberg Astronomical Institute.  SGS acknowledges support through
Grant No. 5-20 of the "Cosmomicrophysics" program of the National
Academy of Sciences of Ukraine to CrAO.  The CrAO CCD cameras have been
purchased through the US Civilian Research and Development Foundation
for the Independent States of the Former Soviet Union (CRDF) awards
UP1-2116 and UP1-2549-CR-03.  This research has made use of the
NASA/IPAC Extragalactic Database (NED) which is operated by the Jet
Propulsion Laboratory, California Institute of Technology, under
contract with the National Aeronautics and Space Administration.


\clearpage


\begin{thebibliography}{77}
\expandafter\ifx\csname natexlab\endcsname\relax\def\natexlab#1{#1}\fi

\bibitem[{{Bennert} {et~al.}(2008){Bennert}, {Canalizo}, {Jungwiert},
  {Stockton}, {Schweizer}, {Peng}, \& {Lacy}}]{Bennert08}
{Bennert}, N., {Canalizo}, G., {Jungwiert}, B., {Stockton}, A., {Schweizer},
  F., {Peng}, C.~Y., \& {Lacy}, M. 2008, \apj, 677, 846

\bibitem[{{Bentz} {et~al.}(2009{\natexlab{a}}){Bentz}, {Peterson}, {Netzer},
  {Pogge}, \& {Vestergaard}}]{Bentz09b}
{Bentz}, M.~C., {Peterson}, B.~M., {Netzer}, H., {Pogge}, R.~W., \&
  {Vestergaard}, M. 2009{\natexlab{a}}, \apj, 697, 160

\bibitem[{{Bentz} {et~al.}(2009{\natexlab{b}}){Bentz}, {Peterson}, {Pogge}, \&
  {Vestergaard}}]{Bentz09a}
{Bentz}, M.~C., {Peterson}, B.~M., {Pogge}, R.~W., \& {Vestergaard}, M.
  2009{\natexlab{b}}, \apjl, 694, L166

\bibitem[{{Bentz} {et~al.}(2006){Bentz}, {Peterson}, {Pogge}, {Vestergaard}, \&
  {Onken}}]{Bentz06a}
{Bentz}, M.~C., {Peterson}, B.~M., {Pogge}, R.~W., {Vestergaard}, M., \&
  {Onken}, C.~A. 2006, \apj, 644, 133

\bibitem[{{Bentz} {et~al.}(2008)}]{Bentz08}
{Bentz}, M.~C., {et~al.} 2008, \apjl, 689, L21

\bibitem[{{Blandford} \& {McKee}(1982)}]{Blandford82}
{Blandford}, R.~D., \& {McKee}, C.~F. 1982, \apj, 255, 419

\bibitem[{{Boller} {et~al.}(1996){Boller}, {Brandt}, \& {Fink}}]{Boller96}
{Boller}, T., {Brandt}, W.~N., \& {Fink}, H. 1996, \aap, 305, 53

\bibitem[{{Chonis} \& {Gaskell}(2008)}]{Chonis08}
{Chonis}, T.~S., \& {Gaskell}, C.~M. 2008, \aj, 135, 264

\bibitem[{{Christopoulou} {et~al.}(1997){Christopoulou}, {Holloway}, {Steffen},
  {Mundell}, {Thean}, {Goudis}, {Meaburn}, \& {Pedlar}}]{Christopoulou97}
{Christopoulou}, P.~E., {Holloway}, A.~J., {Steffen}, W., {Mundell}, C.~G.,
  {Thean}, A.~H.~C., {Goudis}, C.~D., {Meaburn}, J., \& {Pedlar}, A. 1997,
  \mnras, 284, 385

\bibitem[{{Corbett} {et~al.}(2003)}]{Corbett03}
{Corbett}, E.~A., {et~al.} 2003, \mnras, 343, 705

\bibitem[{{Denney} {et~al.}(2009){Denney}, {Peterson}, {Dietrich},
  {Vestergaard}, \& {Bentz}}]{Denney08}
{Denney}, K.~D., {Peterson}, B.~M., {Dietrich}, M., {Vestergaard}, M., \&
  {Bentz}, M.~C. 2009, \apj, 692, 246

\bibitem[{{Denney} {et~al.}(2006)}]{Denney06}
{Denney}, K.~D., {et~al.} 2006, \apj, 653, 152

\bibitem[{{Done} \& {Krolik}(1996)}]{Done96}
{Done}, C., \& {Krolik}, J.~H. 1996, \apj, 463, 144

\bibitem[{{Doroshenko} {et~al.}(2005{\natexlab{a}}){Doroshenko}, {Sergeev},
  {Merkulova}, {Sergeeva}, {Golubinsky}, {Pronik}, \& {Okhmat}}]{Doroshenko05a}
{Doroshenko}, V.~T., {Sergeev}, S.~G., {Merkulova}, N.~I., {Sergeeva}, E.~A.,
  {Golubinsky}, Y.~V., {Pronik}, V.~I., \& {Okhmat}, N.~N. 2005{\natexlab{a}},
  Astrophysics, 48, 156

\bibitem[{{Doroshenko} {et~al.}(2005{\natexlab{b}}){Doroshenko}, {Sergeev},
  {Merkulova}, {Sergeeva}, {Golubinsky}, {Pronik}, \& {Okhmat}}]{Doroshenko05b}
---. 2005{\natexlab{b}}, Astrophysics, 48, 304

\bibitem[{{Edelson} {et~al.}(2002){Edelson}, {Turner}, {Pounds}, {Vaughan},
  {Markowitz}, {Marshall}, {Dobbie}, \& {Warwick}}]{Edelson02}
{Edelson}, R., {Turner}, T.~J., {Pounds}, K., {Vaughan}, S., {Markowitz}, A.,
  {Marshall}, H., {Dobbie}, P., \& {Warwick}, R. 2002, \apj, 568, 610

\bibitem[{{Edelson} \& {Krolik}(1988)}]{Edelson88}
{Edelson}, R.~A., \& {Krolik}, J.~H. 1988, \apj, 333, 646

\bibitem[{{Ferrarese} \& {Merritt}(2000)}]{Ferrarese00}
{Ferrarese}, L., \& {Merritt}, D. 2000, \apjl, 539, L9

\bibitem[{{Ferrarese} {et~al.}(2001){Ferrarese}, {Pogge}, {Peterson},
  {Merritt}, {Wandel}, \& {Joseph}}]{Ferrarese01}
{Ferrarese}, L., {Pogge}, R.~W., {Peterson}, B.~M., {Merritt}, D., {Wandel},
  A., \& {Joseph}, C.~L. 2001, \apjl, 555, L79

\bibitem[{{Fine} {et~al.}(2008)}]{Fine08}
{Fine}, S., {et~al.} 2008, \mnras, 390, 1413

\bibitem[{{Gaskell}(1988)}]{Gaskell88}
{Gaskell}, C.~M. 1988, \apj, 325, 114

\bibitem[{{Gaskell} \& {Peterson}(1987)}]{Gaskell87}
{Gaskell}, C.~M., \& {Peterson}, B.~M. 1987, \apjs, 65, 1

\bibitem[{{Gaskell} \& {Sparke}(1986)}]{Gaskell86}
{Gaskell}, C.~M., \& {Sparke}, L.~S. 1986, \apj, 305, 175

\bibitem[{{Gebhardt} {et~al.}(2000{\natexlab{a}})}]{Gebhardt00a}
{Gebhardt}, K., {et~al.} 2000{\natexlab{a}}, \apjl, 539, L13

\bibitem[{{Gebhardt} {et~al.}(2000{\natexlab{b}})}]{Gebhardt00b}
---. 2000{\natexlab{b}}, \apjl, 543, L5

\bibitem[{{Graham}(2007)}]{Graham07}
{Graham}, A.~W. 2007, \mnras, 379, 711

\bibitem[{{Hopkins} \& {Hernquist}(2009)}]{Hopkins09}
{Hopkins}, P.~F., \& {Hernquist}, L. 2009, \apj, 694, 599

\bibitem[{{Horne} {et~al.}(2004){Horne}, {Peterson}, {Collier}, \&
  {Netzer}}]{Horne04}
{Horne}, K., {Peterson}, B.~M., {Collier}, S.~J., \& {Netzer}, H. 2004, \pasp,
  116, 465

\bibitem[{{Horne} {et~al.}(1991){Horne}, {Welsh}, \& {Peterson}}]{Horne91}
{Horne}, K., {Welsh}, W.~F., \& {Peterson}, B.~M. 1991, \apjl, 367, L5

\bibitem[{{Kaspi} {et~al.}(2005){Kaspi}, {Maoz}, {Netzer}, {Peterson},
  {Vestergaard}, \& {Jannuzi}}]{Kaspi05}
{Kaspi}, S., {Maoz}, D., {Netzer}, H., {Peterson}, B.~M., {Vestergaard}, M., \&
  {Jannuzi}, B.~T. 2005, \apj, 629, 61

\bibitem[{{Kaspi} {et~al.}(2000){Kaspi}, {Smith}, {Netzer}, {Maoz}, {Jannuzi},
  \& {Giveon}}]{Kaspi00}
{Kaspi}, S., {Smith}, P.~S., {Netzer}, H., {Maoz}, D., {Jannuzi}, B.~T., \&
  {Giveon}, U. 2000, \apj, 533, 631

\bibitem[{{Klimek} {et~al.}(2004){Klimek}, {Gaskell}, \& {Hedrick}}]{Klimek04}
{Klimek}, E.~S., {Gaskell}, C.~M., \& {Hedrick}, C.~H. 2004, \apj, 609, 69

\bibitem[{{Kobayashi} {et~al.}(1998{\natexlab{a}}){Kobayashi}, {Yoshii},
  {Peterson}, {Minezaki}, {Enya}, {Suganuma}, \& {Yamamuro}}]{Kobayashi98a}
{Kobayashi}, Y., {Yoshii}, Y., {Peterson}, B.~A., {Minezaki}, T., {Enya}, K.,
  {Suganuma}, M., \& {Yamamuro}, T. 1998{\natexlab{a}}, in Proc. SPIE, Vol.
  3354, 769--776

\bibitem[{{Kobayashi} {et~al.}(1998{\natexlab{b}})}]{Kobayashi98b}
{Kobayashi}, Y., {et~al.} 1998{\natexlab{b}}, in Proc. SPIE, Vol. 3352,
  120--128

\bibitem[{{Kollatschny}(2003)}]{Kollatschny03}
{Kollatschny}, W. 2003, \aap, 407, 461

\bibitem[{{Kollmeier} {et~al.}(2006)}]{Kollmeier06}
{Kollmeier}, J.~A., {et~al.} 2006, \apj, 648, 128

\bibitem[{{Koratkar} \& {Gaskell}(1991)}]{Koratkar91c}
{Koratkar}, A.~P., \& {Gaskell}, C.~M. 1991, \apj, 375, 85

\bibitem[{{Korista} {et~al.}(1995)}]{Korista95}
{Korista}, K.~T., {et~al.} 1995, \apjs, 97, 285

\bibitem[{{Kormendy} \& {Richstone}(1995)}]{Kormendy95}
{Kormendy}, J., \& {Richstone}, D. 1995, \araa, 33, 581

\bibitem[{{Magorrian} {et~al.}(1998)}]{Magorrian98}
{Magorrian}, J., {et~al.} 1998, \aj, 115, 2285

\bibitem[{{Marconi} {et~al.}(2008){Marconi}, {Axon}, {Maiolino}, {Nagao},
  {Pastorini}, {Pietrini}, {Robinson}, \& {Torricelli}}]{Marconi08}
{Marconi}, A., {Axon}, D.~J., {Maiolino}, R., {Nagao}, T., {Pastorini}, G.,
  {Pietrini}, P., {Robinson}, A., \& {Torricelli}, G. 2008, \apj, 678, 693

\bibitem[{{Marconi} {et~al.}(2009){Marconi}, {Axon}, {Maiolino}, {Nagao},
  {Pietrini}, {Risaliti}, {Robinson}, \& {Torricelli}}]{Marconi09}
{Marconi}, A., {Axon}, D.~J., {Maiolino}, R., {Nagao}, T., {Pietrini}, P.,
  {Risaliti}, G., {Robinson}, A., \& {Torricelli}, G. 2009, \apjl, 698, L103

\bibitem[{{Minezaki} {et~al.}(2004){Minezaki}, {Yoshii}, {Kobayashi}, {Enya},
  {Suganuma}, {Tomita}, {Aoki}, \& {Peterson}}]{Minezaki04}
{Minezaki}, T., {Yoshii}, Y., {Kobayashi}, Y., {Enya}, K., {Suganuma}, M.,
  {Tomita}, H., {Aoki}, T., \& {Peterson}, B.~A. 2004, \apjl, 600, L35

\bibitem[{{Nelson} {et~al.}(2004){Nelson}, {Green}, {Bower}, {Gebhardt}, \&
  {Weistrop}}]{Nelson04}
{Nelson}, C.~H., {Green}, R.~F., {Bower}, G., {Gebhardt}, K., \& {Weistrop}, D.
  2004, \apj, 615, 652

\bibitem[{{Nelson} \& {Whittle}(1995)}]{Nelson95}
{Nelson}, C.~H., \& {Whittle}, M. 1995, \apjs, 99, 67

\bibitem[{{Netzer}(2009)}]{Netzer09}
{Netzer}, H. 2009, \apj, 695, 793

\bibitem[{{Netzer} \& {Maoz}(1990)}]{Netzer90b}
{Netzer}, H., \& {Maoz}, D. 1990, \apjl, 365, L5

\bibitem[{{Netzer} {et~al.}(1990){Netzer}, {Maoz}, {Laor}, {Mendelson},
  {Brosch}, {Leibowitz}, {Almoznino}, {Beck}, \& {Mazeh}}]{Netzer90a}
{Netzer}, H., {Maoz}, D., {Laor}, A., {Mendelson}, H., {Brosch}, N.,
  {Leibowitz}, E., {Almoznino}, E., {Beck}, S., \& {Mazeh}, T. 1990, \apj, 353,
  108

\bibitem[{{Onken} {et~al.}(2004){Onken}, {Ferrarese}, {Merritt}, {Peterson},
  {Pogge}, {Vestergaard}, \& {Wandel}}]{Onken04}
{Onken}, C.~A., {Ferrarese}, L., {Merritt}, D., {Peterson}, B.~M., {Pogge},
  R.~W., {Vestergaard}, M., \& {Wandel}, A. 2004, \apj, 615, 645

\bibitem[{{Osterbrock} \& {Pogge}(1985)}]{Osterbrock85}
{Osterbrock}, D.~E., \& {Pogge}, R.~W. 1985, \apj, 297, 166

\bibitem[{{Peterson}(1993)}]{Peterson93}
{Peterson}, B.~M. 1993, \pasp, 105, 247

\bibitem[{{Peterson}(2001)}]{Peterson01}
---. 2001, in Advanced Lectures on the Starburst-AGN Connection, ed. {I.
  Aretxaga, D.} (Singapore: World Scientific), 3

\bibitem[{{Peterson} {et~al.}(1995){Peterson}, {Pogge}, {Wanders}, {Smith}, \&
  {Romanishin}}]{Peterson95}
{Peterson}, B.~M., {Pogge}, R.~W., {Wanders}, I., {Smith}, S.~M., \&
  {Romanishin}, W. 1995, \pasp, 107, 579

\bibitem[{{Peterson} {et~al.}(1998){Peterson}, {Wanders}, {Bertram}, {Hunley},
  {Pogge}, \& {Wagner}}]{Peterson98}
{Peterson}, B.~M., {Wanders}, I., {Bertram}, R., {Hunley}, J.~F., {Pogge},
  R.~W., \& {Wagner}, R.~M. 1998, \apj, 501, 82

\bibitem[{{Peterson} {et~al.}(1991)}]{Peterson91}
{Peterson}, B.~M., {et~al.} 1991, \apj, 368, 119

\bibitem[{{Peterson} {et~al.}(2000)}]{Peterson00b}
---. 2000, \apj, 542, 161

\bibitem[{{Peterson} {et~al.}(2004)}]{Peterson04}
---. 2004, \apj, 613, 682

\bibitem[{{Rodriguez-Pascual} {et~al.}(1997)}]{Rodriguezpascual97}
{Rodriguez-Pascual}, P.~M., {et~al.} 1997, \apjs, 110, 9

\bibitem[{{Russell}(2003)}]{Russell03}
{Russell}, D.~G. 2003, astro-ph/0310284

\bibitem[{{Sergeev} {et~al.}(2005){Sergeev}, {Doroshenko}, {Golubinskiy},
  {Merkulova}, \& {Sergeeva}}]{Sergeev05}
{Sergeev}, S.~G., {Doroshenko}, V.~T., {Golubinskiy}, Y.~V., {Merkulova},
  N.~I., \& {Sergeeva}, E.~A. 2005, \apj, 622, 129

\bibitem[{{Shankar} {et~al.}(2009){Shankar}, {Weinberg}, \&
  {Miralda-Escud{\'e}}}]{Shankar09}
{Shankar}, F., {Weinberg}, D.~H., \& {Miralda-Escud{\'e}}, J. 2009, \apj, 690,
  20

\bibitem[{{Shemmer} {et~al.}(2003){Shemmer}, {Uttley}, {Netzer}, \&
  {McHardy}}]{Shemmer03}
{Shemmer}, O., {Uttley}, P., {Netzer}, H., \& {McHardy}, I.~M. 2003, \mnras,
  343, 1341

\bibitem[{{Shen} {et~al.}(2008{\natexlab{a}}){Shen}, {Vanden Berk},
  {Schneider}, \& {Hall}}]{JShen08}
{Shen}, J., {Vanden Berk}, D.~E., {Schneider}, D.~P., \& {Hall}, P.~B.
  2008{\natexlab{a}}, \aj, 135, 928

\bibitem[{{Shen} {et~al.}(2008{\natexlab{b}}){Shen}, {Greene}, {Strauss},
  {Richards}, \& {Schneider}}]{YShen08}
{Shen}, Y., {Greene}, J.~E., {Strauss}, M.~A., {Richards}, G.~T., \&
  {Schneider}, D.~P. 2008{\natexlab{b}}, \apj, 680, 169

\bibitem[{{Somerville} {et~al.}(2008){Somerville}, {Hopkins}, {Cox},
  {Robertson}, \& {Hernquist}}]{Somerville08}
{Somerville}, R.~S., {Hopkins}, P.~F., {Cox}, T.~J., {Robertson}, B.~E., \&
  {Hernquist}, L. 2008, \mnras, 391, 481

\bibitem[{{Suganuma} {et~al.}(2006)}]{Suganuma06}
{Suganuma}, M., {et~al.} 2006, \apj, 639, 46

\bibitem[{{Tremaine} {et~al.}(2002)}]{Tremaine02}
{Tremaine}, S., {et~al.} 2002, \apj, 574, 740

\bibitem[{{Ulrich} \& {Horne}(1996)}]{Ulrich96}
{Ulrich}, M.-H., \& {Horne}, K. 1996, \mnras, 283, 748

\bibitem[{{van Groningen} \& {Wanders}(1992)}]{vanGroningen92}
{van Groningen}, E., \& {Wanders}, I. 1992, \pasp, 104, 700

\bibitem[{{Vestergaard}(2002)}]{Vestergaard02}
{Vestergaard}, M. 2002, \apj, 571, 733

\bibitem[{{Vestergaard}(2004)}]{Vestergaard04}
---. 2004, \apj, 601, 676

\bibitem[{{Vestergaard} {et~al.}(2008){Vestergaard}, {Fan}, {Tremonti},
  {Osmer}, \& {Richards}}]{Vestergaard08}
{Vestergaard}, M., {Fan}, X., {Tremonti}, C.~A., {Osmer}, P.~S., \& {Richards},
  G.~T. 2008, \apjl, 674, L1

\bibitem[{{Wandel}(2002)}]{Wandel02}
{Wandel}, A. 2002, \apj, 565, 762

\bibitem[{{Welsh} {et~al.}(2007){Welsh}, {Martino}, {Kawaguchi}, \&
  {Kollatschny}}]{Welsh07}
{Welsh}, W.~F., {Martino}, D.~L., {Kawaguchi}, G., \& {Kollatschny}, W. 2007,
  in Astronomical Society of the Pacific Conference Series, Vol. 373, The
  Central Engine of Active Galactic Nuclei, ed. L.~C. {Ho} \& J.-W. {Wang},
  29--+

\bibitem[{{White} \& {Peterson}(1994)}]{White94}
{White}, R.~J., \& {Peterson}, B.~M. 1994, \pasp, 106, 879

\bibitem[{{Yoshii}(2002)}]{Yoshii02}
{Yoshii}, Y. 2002, in New Trends in Theoretical and Observational Cosmology,
  ed. K.~Sato \& T.~Shiromizu (Tokyo: Universal Academy), 235

\bibitem[{{Yoshii} {et~al.}(2003){Yoshii}, {Kobayashi}, \&
  {Minezaki}}]{Yoshii03}
{Yoshii}, Y., {Kobayashi}, Y., \& {Minezaki}, T. 2003, American Astronomical
  Society Meeting Abstracts, 202, 3803

\end{thebibliography}

\clearpage


\begin{deluxetable}{cccc}
\tablecolumns{4}
\tablewidth{0pt}
\tablecaption{$V$-band, Continuum, and H$\beta$ Fluxes for NGC~4051}
\tablehead{
\colhead{JD} &
\colhead{$F_{\lambda}$ (5100 \AA)\tablenotemark{a}} &
\colhead{H$\beta$ $\lambda 4861$} &
\colhead{Observatory}\\
\colhead{($-2450000$)} &
\colhead{($10^{-15}$ erg s$^{-1}$ cm$^{-2}$ \AA$^{-1}$)} &
\colhead{($10^{-13}$ erg s$^{-1}$ cm$^{-2}$)}}

\startdata
4180.30 & 5.03 $\pm$ 0.18   & \nodata           & CrAO \\
4181.37 & 5.66 $\pm$ 0.16   & \nodata           & CrAO \\
4182.02 & 5.47 $\pm$ 0.09   & \nodata           & MAGNUM \\
4182.41 & 5.07 $\pm$ 0.22   & \nodata           & CrAO \\
4184.79 & 5.41 $\pm$ 0.20   & 5.22 $\pm$ 0.11   & MDM \\
4185.71 & 4.51 $\pm$ 0.19   & 4.96 $\pm$ 0.10   & MDM \\
4186.49 & 4.51 $\pm$ 0.22   & \nodata           & CrAO \\
4186.71 & 4.89 $\pm$ 0.20   & 4.15 $\pm$ 0.09   & MDM \\
4187.38 & 4.82 $\pm$ 0.22   & \nodata           & CrAO \\
4187.85 & 5.27 $\pm$ 0.20   & 4.95 $\pm$ 0.10   & MDM \\ 
4188.37 & 4.32 $\pm$ 0.25   & \nodata           & CrAO \\
4188.71 & 4.81 $\pm$ 0.20   & 4.90 $\pm$ 0.10   & MDM \\
4189.39 & 4.75 $\pm$ 0.26   & \nodata           & CrAO \\
4189.61 & 4.78 $\pm$ 0.19   & 4.62 $\pm$ 0.10   & MDM \\
4189.96 & 5.53 $\pm$ 0.21   & 4.88 $\pm$ 0.10   & MDM \\ 
4189.96 & 4.94 $\pm$ 0.05   & \nodata           & MAGNUM\\
4190.41 & 4.82 $\pm$ 0.23   & \nodata           & CrAO \\
4190.72 & 4.82 $\pm$ 0.20   & 4.73 $\pm$ 0.10   & MDM \\ 
4191.38 & 4.59 $\pm$ 0.32   & \nodata           & CrAO \\
4191.62 & 4.91 $\pm$ 0.20   & 4.51 $\pm$ 0.09   & MDM \\
4191.91 & 5.27 $\pm$ 0.20   & 4.77 $\pm$ 0.10   & MDM \\
4192.43 & 4.84 $\pm$ 0.32   & \nodata           & CrAO \\
4192.75 & 4.02 $\pm$ 0.19   & 4.44 $\pm$ 0.09   & MDM \\
4193.61 & 4.86 $\pm$ 0.20   & 4.34 $\pm$ 0.09   & MDM \\
4193.92 & 4.41 $\pm$ 0.19   & 4.57 $\pm$ 0.10   & MDM \\
4194.73 & 4.62 $\pm$ 0.19   & 4.60 $\pm$ 0.10   & MDM \\
4195.45 & 4.49 $\pm$ 0.16   & \nodata           & UNebr. \\
4195.63 & 4.30 $\pm$ 0.19   & 4.06 $\pm$ 0.09   & MDM \\
4196.62 & 4.59 $\pm$ 0.19   & 4.22 $\pm$ 0.09   & MDM \\
4197.78 & 4.22 $\pm$ 0.19   & 4.43 $\pm$ 0.09   & MDM \\
4198.44 & 4.34 $\pm$ 0.14   & \nodata           & UNebr. \\
4198.78 & 4.51 $\pm$ 0.19   & 4.80 $\pm$ 0.10   & MDM \\
4198.93 & 4.46 $\pm$ 0.06   & \nodata           & MAGNUM \\
4199.36 & 4.68 $\pm$ 0.17   & \nodata           & CrAO \\
4199.40 & 4.78 $\pm$ 0.15   & \nodata           & UNebr. \\
4199.86 & 4.55 $\pm$ 0.19   & 4.37 $\pm$ 0.09   & MDM \\
4200.40 & 4.52 $\pm$ 0.15   & \nodata           & CrAO \\
4200.72 & 4.19 $\pm$ 0.19   & 4.29 $\pm$ 0.09   & MDM \\
4201.31 & 4.84 $\pm$ 0.23   & \nodata           & CrAO \\
4201.73 & 4.59 $\pm$ 0.19   & 4.49 $\pm$ 0.09   & MDM \\ 
4202.37 & 4.56 $\pm$ 0.18   & \nodata           & CrAO \\
4202.95 & 4.93 $\pm$ 0.08   & \nodata           & MAGNUM \\
4204.40 & 4.53 $\pm$ 0.17   & \nodata           & CrAO \\
4204.73 & 4.35 $\pm$ 0.19   & 4.58 $\pm$ 0.10   & MDM\\ 
4205.34 & 4.34 $\pm$ 0.16   & \nodata           & CrAO \\
4205.62 & 3.97 $\pm$ 0.18   & 4.37 $\pm$ 0.09   & MDM\\ 
4205.83 & 4.32 $\pm$ 0.03   & \nodata           & MAGNUM \\
4205.91 & 4.28 $\pm$ 0.19   & 4.54 $\pm$ 0.10   & MDM\\ 
4206.36 & 4.19 $\pm$ 0.18   & \nodata           & CrAO \\
4206.40 & 4.23 $\pm$ 0.23   & \nodata           & UNebr. \\
4206.62 & 4.31 $\pm$ 0.19   & 4.46 $\pm$ 0.09   & MDM \\
4207.40 & 4.30 $\pm$ 0.15   & \nodata           & UNebr. \\
4207.82 & 4.40 $\pm$ 0.19   & 4.60 $\pm$ 0.10   & MDM \\
4208.34 & 4.42 $\pm$ 0.15   & \nodata           & CrAO \\
4208.40 & 4.26 $\pm$ 0.17   & \nodata           & UNebr. \\
4208.62 & 4.24 $\pm$ 0.19   & 4.58 $\pm$ 0.10   & MDM \\
4208.88 & 4.25 $\pm$ 0.03   & \nodata           & MAGNUM \\
4209.40 & 4.46 $\pm$ 0.15   & \nodata           & CrAO \\
4209.78 & 4.81 $\pm$ 0.20   & 4.81 $\pm$ 0.10   & MDM \\
4210.62 & 5.01 $\pm$ 0.20   & 4.99 $\pm$ 0.10   & MDM \\
4211.41 & 4.70 $\pm$ 0.31   & \nodata           & CrAO \\
4212.34 & 4.64 $\pm$ 0.15   & \nodata           & CrAO \\
4212.72 & 5.11 $\pm$ 0.20   & 5.01 $\pm$ 0.11   & MDM \\
4212.75 & 4.78 $\pm$ 0.03   & \nodata           & MAGNUM \\
4213.34 & 5.11 $\pm$ 0.19   & \nodata           & CrAO \\
4213.74 & 4.90 $\pm$ 0.20   & 4.50 $\pm$ 0.09   & MDM \\
4214.34 & 4.65 $\pm$ 0.20   & \nodata           & CrAO \\
4214.73 & 4.92 $\pm$ 0.20   & 5.15 $\pm$ 0.11   & MDM \\
4215.40 & 4.72 $\pm$ 0.21   & \nodata           & CrAO \\
4215.74 & 4.84 $\pm$ 0.20   & 4.84 $\pm$ 0.10   & MDM \\
4216.32 & 5.15 $\pm$ 0.21   & \nodata           & CrAO \\
4216.73 & 4.73 $\pm$ 0.19   & 5.09 $\pm$ 0.11   & MDM \\
4217.35 & 4.49 $\pm$ 0.23   & \nodata           & CrAO \\
4217.73 & 4.73 $\pm$ 0.19   & 5.22 $\pm$ 0.11   & MDM \\
4218.31 & 4.58 $\pm$ 0.19   & \nodata           & CrAO \\
4218.80 & 4.31 $\pm$ 0.19   & 4.74 $\pm$ 0.10   & MDM \\
4218.91 & 4.70 $\pm$ 0.04   & \nodata           & MAGNUM \\
4219.32 & 4.73 $\pm$ 0.21   & \nodata           & CrAO \\
4219.40 & 4.71 $\pm$ 0.16   & \nodata           & UNebr. \\
4219.83 & 5.06 $\pm$ 0.20   & 4.70 $\pm$ 0.10   & MDM \\
4220.31 & 4.78 $\pm$ 0.25   & \nodata           & CrAO \\
4220.40 & 5.05 $\pm$ 0.16   & \nodata           & UNebr. \\
4220.74 & 5.18 $\pm$ 0.20   & 4.89 $\pm$ 0.10   & MDM \\
4221.35 & 5.23 $\pm$ 0.38   & \nodata           & CrAO \\
4221.74 & 4.91 $\pm$ 0.20   & 4.75 $\pm$ 0.10   & MDM \\
4221.99 & 4.78 $\pm$ 0.08   & \nodata           & MAGNUM \\
4222.40 & 5.08 $\pm$ 0.26   & \nodata           & CrAO \\
4222.74 & 4.91 $\pm$ 0.20   & 4.83 $\pm$ 0.10   & MDM \\
4223.38 & 4.47 $\pm$ 0.25   & \nodata           & CrAO \\
4223.74 & 4.73 $\pm$ 0.19   & 5.22 $\pm$ 0.11   & MDM \\ 
4224.37 & 5.12 $\pm$ 0.19   & \nodata           & CrAO \\
4224.73 & 4.81 $\pm$ 0.20   & 5.41 $\pm$ 0.11   & MDM \\
4225.35 & 4.46 $\pm$ 0.16   & \nodata           & CrAO \\
4225.80 & 4.30 $\pm$ 0.19   & 4.56 $\pm$ 0.10   & MDM\\ 
4225.88 & 4.43 $\pm$ 0.03   & \nodata           & MAGNUM \\ 
4226.29 & 4.46 $\pm$ 0.20   & \nodata           & CrAO \\
4226.75 & 3.88 $\pm$ 0.18   & 4.55 $\pm$ 0.10   & MDM \\
4227.43 & 4.28 $\pm$ 0.23   & \nodata           & CrAO \\
4227.74 & 4.12 $\pm$ 0.19   & 4.58 $\pm$ 0.10   & MDM \\
4228.80 & 3.81 $\pm$ 0.18   & 4.54 $\pm$ 0.10   & MDM \\
4229.37 & 3.91 $\pm$ 0.15   & \nodata           & CrAO \\
4229.78 & 3.91 $\pm$ 0.18   & 4.31 $\pm$ 0.09   & MDM \\
4230.73 & 4.38 $\pm$ 0.19   & 4.36 $\pm$ 0.09   & MDM \\
4231.36 & 4.07 $\pm$ 0.16   & \nodata           & CrAO \\
4231.59 & 4.58 $\pm$ 0.28   & \nodata           & UNebr. \\
4231.75 & 4.81 $\pm$ 0.20   & 4.61 $\pm$ 0.10   & MDM \\
4232.28 & 4.47 $\pm$ 0.16   & \nodata           & CrAO \\
4232.38 & 4.59 $\pm$ 0.22   & \nodata           & UNebr. \\
4232.73 & 4.22 $\pm$ 0.19   & 4.53 $\pm$ 0.09   & MDM \\
4233.32 & 4.51 $\pm$ 0.15   & \nodata           & CrAO \\
4233.44 & 4.88 $\pm$ 0.18   & \nodata           & UNebr. \\
4233.73 & 4.64 $\pm$ 0.19   & 4.72 $\pm$ 0.10   & MDM \\
4234.32 & 4.17 $\pm$ 0.16   & \nodata           & CrAO \\
4234.73 & 4.51 $\pm$ 0.19   & 4.69 $\pm$ 0.10   & MDM \\
4234.85 & 4.29 $\pm$ 0.03   & \nodata           & MAGNUM \\
4235.31 & 4.35 $\pm$ 0.14   & \nodata           & CrAO \\
4235.46 & 4.80 $\pm$ 0.36   & \nodata           & UNebr. \\ 
4235.73 & 4.43 $\pm$ 0.19   & 4.56 $\pm$ 0.10   & MDM \\
4236.31 & 4.53 $\pm$ 0.15   & \nodata           & CrAO \\
4236.73 & 3.96 $\pm$ 0.18   & 4.54 $\pm$ 0.09   & MDM\\
4237.31 & 4.17 $\pm$ 0.14   & \nodata           & CrAO \\
4237.52 & 4.13 $\pm$ 0.28   & \nodata           & UNebr. \\ 
4237.73 & 3.94 $\pm$ 0.18   & 4.28 $\pm$ 0.09   & MDM \\
4238.49 & 4.38 $\pm$ 0.17   & \nodata           & UNebr. \\
4238.73 & 4.16 $\pm$ 0.19   & 4.42 $\pm$ 0.09   & MDM \\
4238.93 & 4.05 $\pm$ 0.06   & \nodata           & MAGNUM \\
4239.35 & 4.19 $\pm$ 0.17   & \nodata           & CrAO \\
4239.43 & 4.34 $\pm$ 0.18   & \nodata           & UNebr. \\
4239.75 & 3.81 $\pm$ 0.18   & 4.40 $\pm$ 0.09   & MDM \\
4240.29 & 3.69 $\pm$ 0.14   & \nodata           & CrAO \\
4240.47 & 3.69 $\pm$ 0.18   & \nodata           & UNebr. \\
4240.72 & 3.97 $\pm$ 0.18   & 4.37 $\pm$ 0.09   & MDM \\
4241.31 & 3.74 $\pm$ 0.16   & \nodata           & CrAO \\
4241.38 & 3.92 $\pm$ 0.24   & \nodata           & UNebr. \\
4241.73 & 3.75 $\pm$ 0.18   & 4.30 $\pm$ 0.09   & MDM \\
4242.29 & 3.75 $\pm$ 0.14   & \nodata           & CrAO \\
4242.75 & 3.56 $\pm$ 0.18   & 3.91 $\pm$ 0.08   & MDM \\
4243.29 & 3.35 $\pm$ 0.19   & \nodata           & CrAO \\
4243.74 & 3.39 $\pm$ 0.18   & 4.05 $\pm$ 0.09   & MDM \\ 
4244.78 & 3.68 $\pm$ 0.18   & 3.90 $\pm$ 0.08   & MDM \\
4245.35 & 4.38 $\pm$ 0.25   & \nodata           & CrAO \\
4245.75 & 4.36 $\pm$ 0.19   & 4.03 $\pm$ 0.09   & MDM\\ 
4245.77 & 4.21 $\pm$ 0.06   & \nodata           & MAGNUM \\ 
4246.34 & 4.95 $\pm$ 0.18   & \nodata           & CrAO \\
4246.40 & 4.49 $\pm$ 0.17   & \nodata           & UNebr. \\
4246.74 & 4.06 $\pm$ 0.19   & 4.12 $\pm$ 0.09   & MDM \\
4247.74 & 4.57 $\pm$ 0.19   & 4.20 $\pm$ 0.09   & MDM \\
4248.33 & 4.50 $\pm$ 0.21   & \nodata           & CrAO \\
4248.73 & 4.50 $\pm$ 0.19   & 4.48 $\pm$ 0.09   & MDM \\
4249.45 & 4.36 $\pm$ 0.22   & \nodata           & CrAO \\
4249.74 & 4.40 $\pm$ 0.19   & 4.52 $\pm$ 0.09   & MDM \\
4250.41 & 3.56 $\pm$ 0.33   & \nodata           & CrAO \\
4250.74 & 4.07 $\pm$ 0.19   & 4.44 $\pm$ 0.09   & MDM \\
4251.32 & 3.74 $\pm$ 0.30   & \nodata           & CrAO \\
4251.74 & 4.02 $\pm$ 0.19   & 4.39 $\pm$ 0.09   & MDM \\
4252.39 & 4.47 $\pm$ 0.27   & \nodata           & CrAO \\
4252.73 & 4.15 $\pm$ 0.19   & 4.22 $\pm$ 0.09   & MDM \\
4252.88 & 4.17 $\pm$ 0.05   & \nodata           & MAGNUM \\
4253.73 & 4.44 $\pm$ 0.19   & 4.19 $\pm$ 0.09   & MDM \\
4254.35 & 4.62 $\pm$ 0.16   & \nodata           & CrAO \\
4254.39 & 4.52 $\pm$ 0.17   & \nodata           & UNebr. \\
4255.38 & 4.73 $\pm$ 0.24   & \nodata           & CrAO \\
4255.76 & 4.40 $\pm$ 0.19   & 4.50 $\pm$ 0.09   & MDM \\
4256.35 & 5.02 $\pm$ 0.14   & \nodata           & CrAO \\
4256.71 & 4.75 $\pm$ 0.19   & 4.56 $\pm$ 0.10   & MDM \\ 
4257.40 & 4.80 $\pm$ 0.17   & \nodata           & CrAO \\
4257.74 & 4.64 $\pm$ 0.19   & 4.47 $\pm$ 0.09   & MDM \\
4258.35 & 4.52 $\pm$ 0.16   & \nodata           & CrAO \\
4258.50 & 5.06 $\pm$ 0.32   & \nodata           & UNebr. \\ 
4258.76 & 4.83 $\pm$ 0.20   & 4.91 $\pm$ 0.10   & MDM\\ 
4259.34 & 5.15 $\pm$ 0.17   & \nodata           & CrAO \\
4259.75 & 4.92 $\pm$ 0.20   & 4.97 $\pm$ 0.10   & MDM \\
4259.84 & 4.81 $\pm$ 0.13   & \nodata           & MAGNUM \\
4260.30 & 4.96 $\pm$ 0.17   & \nodata           & CrAO \\
4260.75 & 4.36 $\pm$ 0.19   & 4.70 $\pm$ 0.10   & MDM \\
4261.31 & 5.08 $\pm$ 0.16   & \nodata           & CrAO \\
4261.42 & 5.15 $\pm$ 0.14   & \nodata           & UNebr. \\
4261.74 & 4.82 $\pm$ 0.20   & 5.16 $\pm$ 0.11   & MDM \\
4262.30 & 4.78 $\pm$ 0.16   & \nodata           & CrAO \\
4262.74 & 4.39 $\pm$ 0.19   & 4.72 $\pm$ 0.10   & MDM \\
4263.35 & 4.68 $\pm$ 0.17   & \nodata           & CrAO \\
4263.39 & 4.47 $\pm$ 0.22   & \nodata           & UNebr. \\
4263.72 & 4.52 $\pm$ 0.19   & 5.01 $\pm$ 0.11   & MDM \\
4263.88 & 4.61 $\pm$ 0.11   & \nodata           & MAGNUM \\
4264.76 & 4.73 $\pm$ 0.19   & 5.12 $\pm$ 0.11   & MDM \\
4265.76 & 5.43 $\pm$ 0.20   & 5.15 $\pm$ 0.11   & MDM \\
4266.34 & 4.84 $\pm$ 0.27   & 5.18 $\pm$ 0.23   & CrAO \\
4266.76 & 5.55 $\pm$ 0.21   & 5.29 $\pm$ 0.11   & MDM \\ 
4267.30 & 4.43 $\pm$ 0.26   & 5.43 $\pm$ 0.24   & CrAO \\
4267.75 & 4.55 $\pm$ 0.19   & 5.24 $\pm$ 0.11   & MDM\\ 
4268.29 & 4.33 $\pm$ 0.26   & 5.34 $\pm$ 0.23   & CrAO\\ 
4268.75 & 4.38 $\pm$ 0.19   & 5.17 $\pm$ 0.11   & MDM\\ 
4269.32 & 4.57 $\pm$ 0.26   & 5.16 $\pm$ 0.23   & CrAO \\
4269.75 & 4.56 $\pm$ 0.19   & 5.39 $\pm$ 0.11   & MDM \\
4269.84 & 4.37 $\pm$ 0.06   & \nodata           & MAGNUM \\
4270.36 & 5.09 $\pm$ 0.27   & 5.12 $\pm$ 0.23   & CrAO \\
4270.37 & 4.37 $\pm$ 0.20   & \nodata           & UNebr. \\
4271.31 & 4.52 $\pm$ 0.26   & 5.38 $\pm$ 0.24   & CrAO \\
4272.85 & 4.54 $\pm$ 0.04   & \nodata           & MAGNUM \\
4274.31 & 4.78 $\pm$ 0.26   & 5.16 $\pm$ 0.23   & CrAO \\
4275.81 & 4.31 $\pm$ 0.03   & \nodata           & MAGNUM \\
4276.41 & 4.80 $\pm$ 0.20   & \nodata           & UNebr. \\
4277.29 & 4.42 $\pm$ 0.26   & 4.79 $\pm$ 0.21   & CrAO \\
4278.29 & 4.01 $\pm$ 0.25   & 4.99 $\pm$ 0.22   & CrAO \\
4278.48 & 3.66 $\pm$ 0.37   & \nodata           & UNebr. \\
4278.84 & 4.14 $\pm$ 0.06   & \nodata           & MAGNUM \\
4279.31 & 4.33 $\pm$ 0.26   & 4.74 $\pm$ 0.21   & CrAO \\
4279.32 & 4.22 $\pm$ 0.21   & \nodata           & CrAO \\
4280.32 & 4.61 $\pm$ 0.26   & 4.87 $\pm$ 0.21   & CrAO \\
4280.34 & 4.38 $\pm$ 0.43   & \nodata           & CrAO \\
4281.32 & 4.34 $\pm$ 0.24   & \nodata           & CrAO \\
4281.32 & 4.38 $\pm$ 0.26   & 4.88 $\pm$ 0.22   & CrAO \\
4282.33 & 3.96 $\pm$ 0.24   & \nodata           & CrAO \\ 
4282.35 & 4.22 $\pm$ 0.25   & 4.47 $\pm$ 0.20   & CrAO \\
4283.31 & 3.78 $\pm$ 0.25   & 4.63 $\pm$ 0.20   & CrAO \\
4283.33 & 3.75 $\pm$ 0.24   & \nodata           & CrAO \\
4283.42 & 3.56 $\pm$ 0.20   & \nodata           & UNebr. \\
4284.30 & 3.64 $\pm$ 0.24   & 4.50 $\pm$ 0.20   & CrAO \\
4284.31 & 3.87 $\pm$ 0.18   & \nodata           & CrAO \\
4285.77 & 3.70 $\pm$ 0.09   & \nodata           & MAGNUM \\
4289.29 & 4.20 $\pm$ 0.25   & 4.22 $\pm$ 0.19   & CrAO \\
4289.45 & 3.65 $\pm$ 0.25   & \nodata           & UNebr. \\
4290.30 & 3.85 $\pm$ 0.25   & 4.41 $\pm$ 0.19   & CrAO \\
4290.40 & 4.27 $\pm$ 0.22   & \nodata           & UNebr. \\
4291.30 & 3.53 $\pm$ 0.24   & 4.34 $\pm$ 0.19   & CrAO \\ 
4294.29 & 4.63 $\pm$ 0.18   & \nodata           & CrAO \\
4295.32 & 4.51 $\pm$ 0.34   & \nodata           & CrAO \\
4296.29 & 4.35 $\pm$ 0.26   & 4.67 $\pm$ 0.21   & CrAO \\
4296.30 & 4.57 $\pm$ 0.17   & \nodata           & CrAO \\
4298.32 & 4.78 $\pm$ 0.26   & 4.38 $\pm$ 0.19   & CrAO\\ 
4299.28 & 4.29 $\pm$ 0.26   & 4.65 $\pm$ 0.20   & CrAO\\ 
4299.31 & 4.87 $\pm$ 0.16   & \nodata           & CrAO \\
4300.28 & 4.14 $\pm$ 0.25   & 4.68 $\pm$ 0.21   & CrAO\\ 
4304.79 & 4.66 $\pm$ 0.11   & \nodata           & MAGNUM \\
4311.76 & 4.79 $\pm$ 0.09   & \nodata           & MAGNUM
\enddata

\tablenotetext{a}{This column contains the average continuum flux
density measured at rest-frame $\sim$5100\AA\ from spectroscopic
observations as well as the $V$-band flux from photometric observations.
Spectroscopic and photometric fluxes were intercalibrated and merged to
create a single continuum light curve (see Section 2.3).}

\label{tab:fluxes}
\end{deluxetable}

\begin{deluxetable}{cccccccc}
\tablecolumns{8}
\tablewidth{0pt}
\tablecaption{Light Curve Statistics}
\tablehead{
\colhead{ }&\colhead{ }&\multicolumn{2}{c}{Sampling}&\colhead{ }&
\colhead{Mean}&\colhead{ }&\colhead{ }\\
\colhead{Time}&\colhead{ }&\multicolumn{2}{c}{Interval(days)}&\colhead{Mean}&
\colhead{Fractional}&\colhead{ }&\colhead{ }\\
\colhead{Series}&\colhead{$N$}&\colhead{$\langle T \rangle$}&
\colhead{$T_{\rm median}$}&\colhead{Flux\tablenotemark{a}}&\colhead{Error}&
\colhead{$F_{\rm var}$}&\colhead{$R_{\rm max}$}\\
\colhead{(1)}&\colhead{(2)}&\colhead{(3)}&\colhead{(4)}&\colhead{(5)}&
\colhead{(6)}&\colhead{(7)}&\colhead{(8)}}
\startdata

$5100$ \AA & $186$ & $0.71$ & $0.56$ & $4.5 \pm 0.4$ & $0.04$ & $0.09$ & $1.69 \pm 0.11$\\
 H$\beta$ & $100$ & $1.17$ & $1.00$ & $4.7 \pm 0.3$ & $0.02$ & $0.07$ & $1.39 \pm 0.04$\\


\enddata
\tablenotetext{a}{Same flux units as Table 1 for $5100$ \AA\ continuum and H$\beta$, respectively.}
\label{tab:lcstats}
\end{deluxetable}

\clearpage

\begin{deluxetable}{cc}
\tablecolumns{2}
\tablewidth{0pt}
\tablecaption{Reverberation Results}
\tablehead{
\colhead{Parameter}&{Value}\\
\colhead{(1)}&{(2)}}
\startdata
$\tau_{\rm cent}$ & $1.87^{+0.54}_{-0.50}$ days\\
$\tau_{\rm peak}$ & $2.60^{+0.79}_{-1.40}$ days\\
$\sigma_{\rm line} (\rm mean)$ & $1045 \pm 4$ km s$^{-1}$\\
FWHM (mean)& $799 \pm 2$ km s$^{-1}$\\
$\sigma_{\rm line} (\rm rms)$ & $927 \pm 64$ km s$^{-1}$\\
FWHM (rms)& $1034 \pm 41$ km s$^{-1}$\\
$M_{\rm BH}$\tablenotemark{a} & $(1.73^{+0.55}_{-0.52}) \times 10^{6}M_{\odot}$\\
$M_{\rm BH-rad}$\tablenotemark{b} & $(1.24^{+0.57}_{-0.56}) \times 10^{6}M_{\odot}$\\
\enddata
\tablenotetext{a}{Using \citet{Onken04} calibration.}
\tablenotetext{b}{Using \citet{Marconi08} calibration.}

\label{tab:results}
\end{deluxetable}

\clearpage


\end{document}